\documentclass{article}

\usepackage{microtype}
\usepackage{graphicx}
\usepackage{subfigure}
\usepackage{booktabs} % for professional tables
\usepackage{color}          % text color
\usepackage{bm}
\usepackage{multirow, booktabs}
\definecolor{Graylight}{gray}{0.9}
\usepackage{comment}
\usepackage{xcolor,colortbl}
\usepackage{makecell}

\usepackage{hyperref}

\usepackage[accepted]{icml2023}

\usepackage{amsmath}
\usepackage{amssymb}
\usepackage{mathtools}
\usepackage{algorithmic}

\usepackage[capitalize,noabbrev]{cleveref}
\usepackage[textsize=tiny]{todonotes}

\usepackage{qiangstyle}

\newcommand{\lion}{Lion}
\newcommand{\mavolion}{Distributed Lion}

\newcommand{\adam}{Adam}
\newcommand{\adamw}{AdamW}

\theoremstyle{plain}

\theoremstyle{definition}

\theoremstyle{remark}

\definecolor{myblue}{HTML}{0153d6} % Replace 0000FF with the hex code for your blue
\definecolor{myred}{HTML}{ff409b} % Replace FF0000 with the hex code for your red

\newcommand{\myobs}[1]{
  \vspace{5pt}\noindent\textbf{\textcolor{myblue}{#1}}
}

\icmltitlerunning{Communication Efficient Distributed Training with Distributed Lion}

\begin{document}

\twocolumn[
\icmltitle{Communication Efficient Distributed Training with Distributed Lion}
% \icmltitle{Distributed Lion: An Efficient Approach to Low Bandwidth Distributed Training}



\icmlsetsymbol{equal}{*}

\begin{icmlauthorlist}
\icmlauthor{Bo Liu}{equal,ut}
\icmlauthor{Lemeng Wu}{equal,ut,comp}
\icmlauthor{Lizhang Chen}{equal,ut}
\icmlauthor{Kaizhao Liang}{ut}
\icmlauthor{Jiaxu Zhu}{comp}
\icmlauthor{Chen Liang}{}
\icmlauthor{Raghuraman Krishnamoorthi}{comp}
\icmlauthor{Qiang Liu}{ut}
\end{icmlauthorlist}

\icmlaffiliation{ut}{The University of Texas at Austin}
\icmlaffiliation{comp}{Meta Reality Lab}

\icmlcorrespondingauthor{Bo Liu}{bliu@cs.utexas.edu}
%feel free to add
\icmlcorrespondingauthor{Lizhang Chen}{lzchen@utexas.edu}

\icmlkeywords{Machine Learning, ICML}

\vskip 0.3in
]

\printAffiliationsAndNotice{\icmlEqualContribution} % otherwise use the standard text.

\begin{abstract}
The \lion{} optimizer 
has been a promising competitor with the \adamw{} 
for training large AI models,
with advantages on memory, computation, and sample efficiency. 
In this paper, we introduce \mavolion{}, an innovative adaptation of \lion{} for distributed training environments. 
Leveraging the sign operator in \lion{}, 
our \mavolion{}  
only requires to 
communicate binary or lower-precision vectors
between workers to the center server, 
significantly reducing the communication cost.  
Our theoretical analysis confirms \mavolion{}'s convergence properties. Empirical results demonstrate its robustness across a range of tasks, worker counts, and batch sizes, on both vision and language problems. Notably, \mavolion{} attains comparable performance to standard \lion{} or \adamw{} optimizers applied on aggregated gradients, but with significantly reduced communication bandwidth. This feature is particularly advantageous for training large models. In addition, we also demonstrate that \mavolion{} presents a more favorable performance-bandwidth balance compared to existing efficient distributed methods such as deep gradient compression and ternary gradients.
\end{abstract}

\section{Introduction}
\label{sec::intro}
The pursuit of modern artificial intelligence hinges on the training of large-scale models like large language models\citep{OpenAI2023GPT4TR} and large vision models (LVM)\citep{kirillov2023segment}. As the stakes – in terms of time, cost, and environmental impact – grow ever higher for training expansive AI systems, the hunt for efficient optimizers becomes critical.

Recently, a new optimization named \lion{} (evolved sign momentum)~\citep{chen2023symbolic} has been discovered with an evolutionary program. 
It was shown that it exhibits performance on par with the current state-of-the-art AdamW~\citep{loshchilov2017decoupled} across a wide range of tasks, while reducing the memory cost and training time. 

Consider optimizing a loss function $f_{\mathcal D}(x)$ on $ \RR^d$ associated with a dataset $\mathcal{D}$, the update rule of Lion is:
\begin{equation}
\label{equ:lion-update} 
\begin{split}
&m_{t+1}  = \btwo m_t + (1-\btwo)\dd f_{\mathcal D}(x_t), \\ 
&\delta_t = \texttt{Lion}(x_t, \mathcal{D}) \overset{def}{=} \sign(\bone m_t + (1-\bone)\dd f_{\mathcal D}(x_t)), \\
&x_{t+1}  = x_t - \lr \big(\delta_t + \lambda x_t\big),
\end{split}
\end{equation}
where $m_t$ plays the role of the momentum, $\lr$ is the learning rate,  $\bone, \btwo \in [0,1]$\footnote{\citet{chen2023symbolic} suggests $(\bone=0.9, \btwo=0.99)$ based on empirical findings.} are two momentum related coefficients, and $\lambda \geq 0$ is the weight decay coefficient. Comparing \lion{} against \adamw{}, one observes that \lion{} only requires the storage of the first-order momentum term, which results in a more relaxed memory requirement.

\begin{figure}[t!]
    \centering
    \includegraphics[width=0.9\columnwidth]{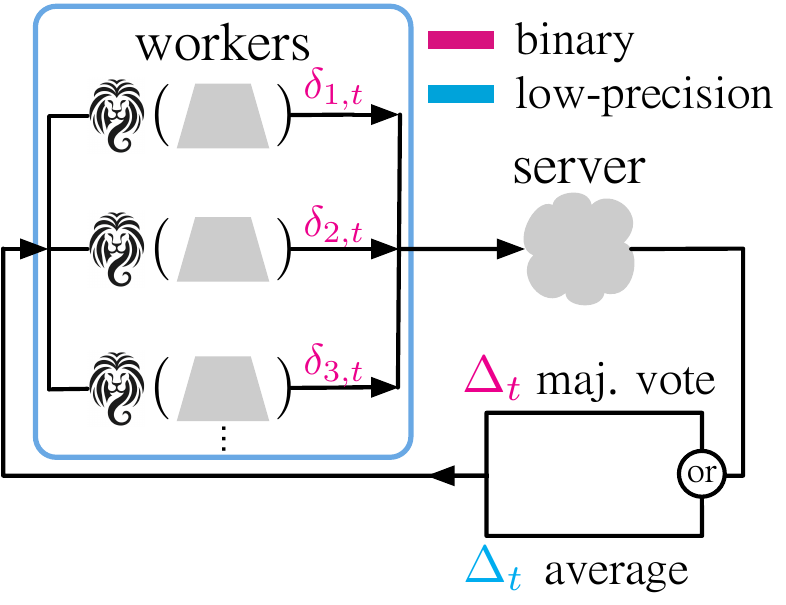}
    \caption{Illustration of Distributed-Lion. Each worker keeps its \emph{own} optimizer state and applies the \lion{} optimizer individually to a binary update \textcolor{magenta}{$\delta_{i,t} =\texttt{Lion}(x, \mathcal D_i)$} (without the weight decay), then the server aggregates all \textcolor{magenta}{$\delta_{i,t}$} to produce 
 a binary \textcolor{magenta}{$\Delta_t$} by majority vote (or a integer \textcolor{cyan}{$\Delta_t$} by averaging) and send it back to all workers. The workers then apply $\Delta_t$ and weight decay to update their model parameters. See  Algorithm~\ref{alg:dist-lion} for details.}
    \label{fig:intro}
\end{figure}

In this study, 
we tailor the \lion{} optimizer for distributed training. The \lion{} optimizer is particularly suitable for this context due to two main attributes: (1) its simple update mechanism that relies solely on first-order momentum, and (2) its use of the $\sign(\cdot)$ function. We showcase the effective employment of the $\sign(\cdot)$ function to streamline communication processes, leading to the development of a novel distributed training framework named \mavolion{}. Within the \mavolion{} framework, each participating worker independently adjusts the model parameters using a distinct instance of the \lion{} optimizer, thereby maintaining separate optimizer states. A distinctive feature of this framework is the mode of communication between workers and the central server, which is restricted to binary or low-precision vectors.

Crucially, in this setup, workers convey updates rather than raw gradients to the central server. The server, in turn, aggregates these updates through either a straightforward averaging process (\mavolion{}-Avg) or a majority voting mechanism (\mavolion{}-MaVo). In the case of \mavolion{}-MaVo, the consolidated update is maintained as a binary vector, whereas for \mavolion{}-Avg, given the presence of $n$ workers, each element of the update vector is encoded using $\log(n)$ bits. This approach markedly reduces the bandwidth requirements compared to traditional distributed training methods, which typically rely on high-precision floating-point vectors for communication. The bandwidth efficiencies achieved by our method are detailed in Table~\ref{tab:bandwidth}. We summarize our primary contributions as follows:

\begin{table}[t!]
\centering
 \renewcommand\arraystretch{1.3}
\resizebox{\linewidth}{!}{
\begin{tabular}{l c c}
\toprule
& \multicolumn{2}{c}{Bandwidth Requirement} \\
\cmidrule{2-3}
\multirow{-2}{*}{\centering Method} & Worker$\rightarrow$Server & Server$\rightarrow$Worker \\
\midrule
Global Lion/AdamW & $32d$ & $32d$ \\
TernGrad~\citep{wen2017terngrad} & $1.5d$ & $\log(2n+1)d$ \\
DGC~\citep{lin2017deep} & $(1 -\eta) 32d$ & $32d$ \\
\midrule
\mavolion{}-Avg & $d$ & $\log(n)d$ \\
\mavolion{}-MaVo & $d$ & $d$ \\
\bottomrule
\end{tabular}
}
\caption{Minimum bandwidth requirements of different methods for a model with $d$ parameters and $n$ workers. For Deep Gradient Compression (DGC), $\eta$ denotes the compression rate (default: $\eta=0.96$).}
\label{tab:bandwidth}
\end{table}

\begin{itemize}
\item We introduce the \mavolion{} algorithm, a simple yet effective approach to extend \lion{} to distributed training, where all communications between workers and the server are done through binary or low-precision vectors (Section~\ref{sec::method}).
\item We provide theoretical analysis to ensure the convergence of \mavolion{} (Section~\ref{sec::theory}).
\item Empirically, we demonstrate that on both vision and language modeling tasks, \mavolion{} achieves comparable performance against applying \lion{} and \adam{} with the synchronized gradients from all workers, while being significantly more communication efficient. In addition, we show that \mavolion{} achieves a better trade-off than existing efficient distributed training methods like deep gradient compression~\citep{lin2017deep} and ternary gradients~\citep{wen2017terngrad} (Section~\ref{sec::exp}).
\end{itemize}

\section{Related Work}
\label{sec::related}
In this section, we provide a summary of optimizers that use the sign function and existing literature on bandwidth-friendly distributed training.

\paragraph{Sign Operation in Optimization}
The sign operation is integral to optimization for several reasons. Primarily, it acts as a normalization mechanism by disregarding the magnitude of gradients, thereby equilibrating updates across different dimensions and potentially facilitating the avoidance of saddle points. Additionally, the binary nature of the sign function's output significantly reduces the memory footprint required for storing gradient updates. The concept of sign-based optimization dates back to RProp~\citep{riedmiller1993direct} and has seen renewed interest with the advent of SignSGD and its momentum-enhanced variant, Signum~\citep{bernstein_signsgd_2018}. A more recent advancement is the generalized SignSGD algorithm introduced by \cite{crawshaw_robustness_2022}, which incorporates a preconditioner, making it a superset of SignSGD and akin to Adam in certain aspects. A noteworthy addition to sign-based optimizers is the Lion optimizer, which emerged from evolutionary program search, achieving performance comparable to Adam~\citep{kingma2014adam} and AdamW~\citep{loshchilov2017decoupled} for the first time. Lion distinguishes itself from Signum by employing a different convex combination for outputting local updates, a technique referred to as the double-$\beta$ scheme, reminiscent of Nesterov's momentum update, and encapsulates Signum as a particular case. On the theoretical front, SignSGD and Signum have been shown to exhibit convergence rates comparable to traditional SGD~\citep{bernstein_signsgd_2018}. Recent work by \cite{sun2023momentum} has extended the theoretical understanding by providing a convergence theory that relaxes the requirements for bounded stochastic gradients and enlarged batch sizes. Additionally, Lion has demonstrated its capability in performing constrained optimization under the $\ell_\infty$-norm constraint~\citep{chen2023lion}.

\paragraph{Distributed Training}
In addressing the communication constraints of distributed training, the research community has devised several innovative strategies, prominently featuring asynchronous Stochastic Gradient Descent (SGD), gradient quantization, and sparsification techniques. Asynchronous SGD offers a solution by enabling parameter updates immediately after back-propagation, bypassing the need for gradient synchronization, thereby expediting the training process~\cite{chen2016revisiting, zheng2017asynchronous, liu2024asynchronous}.  \citet{li2022stingy} utilizes sketch-based algorithms for lossless data compression \cite{li2024accelerating}, achieving an asymptotically optimal compression ratio \cite{li2023chainedfilter}. However, its applicability is limited to highly sparse gradients, making it orthogonal to our research. In the realm of gradient quantization, methods such as 1-bit SGD~\cite{seide20141}, QSGD~\cite{alistarh2017qsgd}, and TernGrad~\cite{wen2017terngrad} are pivotal. These approaches compact the gradient data, substantially reducing the required communication bandwidth, with 1-bit SGD demonstrating a tenfold acceleration in speech applications and both QSGD and TernGrad confirming the feasibility of quantized training in maintaining convergence. Moreover, gradient sparsification further mitigates the communication load by transmitting only the most substantial gradients. Techniques like threshold quantization and Gradient Dropping~\cite{aji2017sparse} exemplify this, with Gradient Dropping notably achieving a 99 reduction in gradient exchange with minimal impact on performance metrics, such as a mere 0.3 loss in BLEU score for machine translation tasks. The recent Deep Gradient Compression (DGC) strategy~\cite{lin2017deep} also contributes to this field by incorporating momentum correction and local gradient clipping among other methods to maintain accuracy while significantly reducing communication demands, albeit at the cost of increased computational overhead. Compared to gradient quantization methods, \mavolion{} uniquely leverages the binary nature of Lion's update and can be viewed as performing quantization on updates rather than the gradient.
\begin{algorithm*}[t!]
\caption{\mavolion{} Training}
\begin{algorithmic}
    \STATE\textbf{Inputs:} Initial parameters $x_0 \in \mathbb{R}^d$, datasets $\{\mathcal{D}_1, \dots, \mathcal{D}_N\}$, loss function $f$, learning rate $\lr$, hyper-parameters $\bone, \btwo \in [0, 1]$ (default to $0.9, 0.99$)\footnote{We allow each worker to set its individual $\bone, \btwo$.}, and the weight decay $\lambda$.\\
    \STATE

    \STATE \textbf{Initialization:} $t = 0$, ~$\forall i, m_{i,0} = \bm{0}$, and $x_{i,0} = x_0$.
    \WHILE{not convergent}{
        \STATE \textbf{Worker-side:} Each worker $i$ samples a batch $\xi_{i,t} \in D_i$, computes the following, and sends $\textcolor{magenta}{\delta_{i,t}}$ to the server:
            \begin{flalign*}
            \text{if}~t > 0,~~x_{i,t} &\leftarrow x_{i,t-1} - \epsilon\big(\Delta_{t-1} +\lambda x_{i,t-1} \big)\\
            \textcolor{magenta}{\delta_{i,t}} &\leftarrow \sign\big(\bone m_{i,t} + (1 - \bone)\nabla_x f(x_{i,t}; \xi_{i,t})\big)\\
            m_{i,t+1} &\leftarrow \btwo m_{i,t} + (1 - \btwo) \nabla_x f(x_{i,t}; \xi_{i,t}).
            \end{flalign*}\\
        \STATE \textbf{Server-side:} The server computes the aggregated update $\Delta_t$ and broadcast it to all workers:
        $$
        \Delta_t = \begin{cases}
            \textcolor{cyan}{\frac{1}{N} \big(\sum_{i=1}^N \delta_{i,t}\big)} & \text{(Averaging)} \\
            \textcolor{magenta}{\sign \big(\sum_{i=1}^N \delta_{i,t}\big)} & \text{(Majority Vote)} \\
        \end{cases} 
        ~~~~\text{and}~~~~t \leftarrow t + 1.
        $$
        }
    \ENDWHILE
\end{algorithmic}
\label{alg:dist-lion}
\end{algorithm*}
% \vspace{-10pt}

\section{The \mavolion{}}
\label{sec::method}
We  introduce the distributed learning problem and then our \mavolion{} framework. 
%In this section, we formally present the distributed training problem, then introduce the relevant notations and the \mavolion{} algorithm.

\subsection{Distributed Training} In distributed training, we aim to minimize the following learning objective:
\begin{equation}
\label{equ:distibuted-objective}
\min_x F(x) = \frac{1}{N} \sum_{i=1}^N \mathbb{E}_{\xi_i \sim \mathcal{D}_i} \bigg[ f(x; \xi_i)\bigg].
\end{equation} 
Here, $N$ denotes the number of workers, $\{\mathcal{D}_i\}$ are $N$ datasets,\footnote{Throughout this work, we assume $\{\mathcal{D}_i\}$ consist of i.i.d data samples, $\xi_i$ sampled from $\mathcal{D}_i$ is i.i.d. though our method should be directly applicable to non-i.i.d data.} and $x$ is the model parameter (e.g., the weights of a neural network). In the distributed learning setting, each worker $i \in [n]$ will get its own dataset $\mathcal{D}_i$, and we assume there is a centralized server that all workers can communicate with. The simplest distributed training technique is to perform distributed gradient aggregation:
\begin{equation}
    g_\text{server} = \frac{1}{N} \sum_{i=1}^N g_i,~~~\text{where}~~~g_i = \mathbb{E}_{\xi_i \sim \mathcal{D}_i} \big[ \nabla_x f(x; \xi_i)\big].
\end{equation}
Here, each local gradient $g_i$ is an unbiased estimation of the true gradient $\nabla_x F(x)$ when $\mathcal D_i$ are i.i.d. drawn from the same underlying distribution. The server aggregates all local gradients into $g_\text{server}$, and then applies an optimizer like Adam~\citep{kingma2014adam} on top of $g_\text{server}$. However, 
the aggregation step requires communicating the full gradient vectors $g_i$, 
which can be expensive for large models.

\paragraph{Notation.} Given a function $f(x;\xi)$, the gradient $\dd f(x;\xi)$ is taken with respect to variable $x$. We use $\|\cdot\|$, $\|\cdot\|_1$, and $\|\cdot\|_\infty$ to denote the $\ell_2$, $\ell_1$, and $\ell_\infty$ norm, respectively. $\xi_{i,t}$ is the sampled data at time $t$ for the $i$-th worker and $g_{i,t} = \dd f(x_t;,\xi_{i,t})$. We similarly denote $z_{i,t}$ as any variable $z$ at time $t$ from worker $i$.

\subsection{\mavolion{}}
The main idea of \mavolion{} is to leverage the binary nature of the \lion{}'s update for efficient communication. To enable that, we want the workers to \emph{only send the binary updates}  to the server. As a result, we let each worker keep tracks of its own optimizer state, i.e., the momentum $m_{i,t}$. Then at each step, each worker $i$ first computes:
\begin{equation}
\begin{split}
m_{i, t+1} = \beta_2 m_{i, t} + (1 - \beta_2) g_{i, t}, \\
\textcolor{magenta}{\delta_{i, t}} = \sign(\beta_1 m_{i, t} + (1 - \beta_1) g_{i, t}). \\
\end{split}
\end{equation}
Then all workers send the \textcolor{magenta}{$\delta_{i, t}$} back to the server. The server receives the binary ``updates" from all workers and then aggregates them. Here, we propose two simple ways for aggregation. Denote $S_t = \sum_{i=1}^N \delta_{i, t}$, which is a vector of integers in $\{0,\ldots N\}$. Define the  aggregation as follows: 
\begin{equation}
\Delta_t = \text{aggregate}(S_t) = \begin{cases}
    \textcolor{cyan}{\frac{1}{N} S_t} & \text{(Averaging)}\\
    \textcolor{magenta}{\sign(S_t)} & \text{(Majority Vote)}\\
\end{cases}.
\end{equation}
So we simply average or take the majority vote from all $\{\delta_{i, t}\}$. Here, we denote binary vectors in \textcolor{magenta}{magenta} and low precision vectors in \textcolor{cyan}{cyan}. In the end, the server broadcasts $\Delta_t$ back to each worker $i$, and each worker performs
\begin{equation}
x_{i, t+1} = x_{i, t} - \epsilon(\Delta_t + \lambda x_{i, t}),
\end{equation}
where $\epsilon$ is the step size and $\lambda$ is the weight decay coefficient. 

\paragraph{Communication Cost}
In both variants of \mavolion{}, 
the $N$ workers only need to send the binary 
vectors \textcolor{magenta}{$\delta_{i,t}$} to the server. 
The servers need to send the aggregated updates $\Delta_t$ back to the workers, 
which is binary when using the majority vote aggregation, and an integer in $\{0, \ldots, N\}$ 
when using the averaging aggregation. 
Note that an integer in $\{0,\ldots, N\}$ 
can be represented by at most $\log(N)$ bits. 
In practice, usually $N \ll 2^{32}$ hence $\log(N) < 32$ and we still save the communication bandwidth even with the average aggregation, comparing against communicating with floating point numbers (Check Table~\ref{tab:bandwidth}). The whole \mavolion{} algorithm is summarized in Algorithm~\ref{alg:dist-lion}.

\section{Theoretical Analysis}
\label{sec::theory}
We provide our theoretical analysis of the \mavolion{} algorithm, both with the averaging and the majority vote aggregation methods. In the following, we first describe that the distributed training problem can be viewed as a constrained optimization problem when \mavolion{} is used. We provide convergence results for \mavolion{} with both aggregation methods.

\subsection{Lion as Constrained Optimization}
\label{sec::theory-constrained}

\citet{chen2023lion} showed that the (global) Lion is a theoretically novel and principled approach for minimizing a general loss function $f(x)$ while enforcing a box constrained optimization problem:  %$\norm{x}_\infty \leq1/\lambda$: % Lion achieves this through the incorporation of decoupled weight decay, where $\lambda$ represents the weight decay coefficient. 
\bbb \label{equ:lionboundc}
\min_{x\in\RR^d} f(x)~~~~s.t.~~~~ \norm{\lambda x}_\infty \leq 1, 
\eee 
where the constrained is introduced 
due to the use of %where the bound $1/\lambda$ is solely decided by 
the weight decay coefficient $\lambda$. 

Moreover, \citet{chen2023lion} showed that the Lion dynamics consists of two phases: 

1) \textbf{[Phase 1]} When 
the constraint is not satisfied, that is, $x \not\in\mathcal F $, where $\mathcal F$ is the feasible set 
\bbb  
\mathcal F \overset{def}{=} \{x \colon \norm{\lambda x }_\infty \leq 1\},
\eee  
% \qq{define $\mathcal F$ so we do not need to repeat the definition later}
%$\n%orm{\lambda x_t}_\infty > 1$,  
it exponentially decays the distance to $\mathcal F$: 
%from $\lambda x_t$ to the set 
% \qq{since we did not introduce the continuous form of Lion, we need to recall the discrete version of the bound here.}
there exists an $\alpha \in (0,1)$, such that 
% $$
% \dist(x_t, \mathcal F) \leq \exp(-\lambda (t-s) ) ~\dist(x_s, \mathcal F),~~~\forall s \leq t. 
% $$
% \qq{use this one}\lzchen{removed the continuous bound, just used the following one}
$$
\dist(x_{t+n}, \mathcal{F}) \leq \alpha^{n} \dist(x_t, \mathcal{F}).
$$
where $n\geq 0$. 
% \qq{if we introduce the implicit form, we can use $\alpha = \frac{1}{1+\epsilon \lambda }$ here. Note the slight gap if we do not use the implicit form here.}\lzchen{here we should introduce both the implicit and explicit forms, or we can just say "there exits an $\alpha$" to keep it simple and concise.}
%where $0<\alpha <1$ is a 
Hence, $x_t$ converges to $\mathcal F$ rapidly and stays within $\mathcal F$ once it arrived it.  

%Figure~\red{xxx} shows that the bound constraint is enforced well in large vision and language models. In addition, the constraint is enforced rapidly when violated 
%\red{(e.g., xx steps in xx model)}; this is because when $\abs{x_i}$ is large, the weight decay term $\lambda x_i$ dominates the signed momentum term, and decreases $\abs{x_i}$ exponentially fast.  

2) \textbf{[Phase 2]} 
After $\lambda x_t$ enters $\mathcal F$, the dynamics minimizes the objective $f(x)$ while being confined within the set $\mathcal F$. 
This step is proved in \citet{chen2023lion}
by constructing a Lyapunov function when $\sign(\cdot)$ 
is treated as the sub-gradient of a convex function. 
%the finite valued objective $F(x)$. This is proved by showing that the 
%Lion-$\phi$ dynamics minimizes the following Lyapunov function: 
\iffalse 
\bbb  \label{equ:H1}
H(x, m) = \alpha   f(x)   + 
\frac{ \gamma }{\lambda}
\phi^*(\lambda x) + 
\frac{1-\varepsilon \gamma }{1+ \varepsilon \lambda} 
(\phi^*(\lambda x) +  \phi(m)   - \lambda 
m\tt  x ). 
\eee
We show that, whenever $H(x_t,m_t)$ is finite, it is 
decreased monotonically (i.e., $\ddt H(x_t, m_t) \leq 0$) along trajectories of \eqref{equ:lionode} until a local minimum of point of $H(x,m)$ is reached.  
\fi

%In distributed training scenarios like Majority Voting, Averaging, or Global Lion, the Lion optimizer adeptly solves a constrained optimization problem. This capability underscores Lion's adaptability across various training setups, maintaining its core optimization strategy of minimizing a general loss function within a specific bound constraint.

\subsection{Convergence Analysis}
In this section, 
we analyze the convergence of distributed Lion algorithms.
Similar to the case of global Lion, 
 we show that distributed Lion also solves the box constrained optimization \eqref{equ:lionboundc}. 
 Its dynamics also unfolds into two phases 
%The analysis unfolds in two parts, 
aligning with Lion's dynamics: Phase I shows rapid convergence to a feasible set $\mathcal F$, while Phase II seeks to minize the objective $f(x)$ within the feasible set $\mathcal F$. 
Different from the Lyapunov approach used in \citet{chen2023lion},  
the proof of our Phase II result 
is made by introducing 
a surrogate metric \(\mathcal{S}(x)\) of constrained optimality, and providing upper bound of \(\mathcal{S}(x_t) \) following the algorithm. 
%indicating a stationary solution. 
%Theorems for Majority Vote, Global Lion, and Averaging demonstrate effective convergence rates.

Our analysis makes the following assumptions. 
%We first list the assumptions we make for the analysis in the following.
% \begin{ass}[Unbiased Stochastic Gradient]\label{ass::sample}
%     Given a dataset $\mathcal{D}$, the stochastic sample $\xi \sim \mathcal{D}$ is i.i.d. and 
%     \bb 
%     &\E[\dd f(x, \xi)] = \dd f(x), \\
%     &\E\left[\|\dd f(x;\xi)-\dd f(x)\|^2\right]  \leq \sigma^2.
%     \ee
% \end{ass}
\begin{ass}[Variance bound]\label{ass::sample} $\mathcal{D}_i$ is i.i.d. drawn from a common distribution $\mathcal \pi_*$, and the stochastic sample $\xi^i \sim \mathcal{D}_i$ is i.i.d. and upon receiving query $x \in \mathbb{R}^d$, the stochastic gradient oracle gives us an \emph{independent} unbiased estimate $\dd f(x;\xi^i)$ from the  $i$-th worker that has coordinate bounded variance:
\bb
&\mathbb{E}_{\xi}[\dd f(x;\xi^i)]= \dd f(x),\\
&\E_\xi\left[\|\dd f(x;\xi^i)-\dd f(x)\|^2\right]  \leq \sigma^2.
\ee
\end{ass}
\begin{ass}[Smooth and Differentiable $f$]\label{ass::L-smooth}
    Function $f(\cdot)$ is differentiable and L-smooth. 
\end{ass}

% \begin{ass}[Bias Correction]\label{ass::bias_corr}
%  Consider the sequence $\{m_{i,t}\}_{t>0,i \in [N]}$ generated by Algorithm \ref{alg:dist-lion}. We define the ratio $R_t$ as $ \E [\beta_1 m_{i, t} + (1 - \beta_1) g_{i, t}] / \E [\delta_{i,t}]$, where $R_t$ is non-negative.
% \end{ass}

\begin{ass}[Bias Correction]\label{ass::bias_corr}
  Consider the sequence $\{m_t^i\}_{t>0,i \in [N]}$ generated by Algorithm \ref{alg:dist-lion}, $\E [\tilde m_{t}^i] / \E [\sign(\tilde m_{t}^i)] \geq 0$.
\end{ass}

Note that assumption \ref{ass::L-smooth} \ref{ass::sample} are standard in the analysis of stochastic optimization algorithms~\cite{bottou2018optimization, sun2023momentum}. 
When Assumption~\ref{ass::sample} holds, $\E \|\frac{1}{N}\sum_{i=1}^N\dd f(x;\xi_i) - \dd f(x)\|^2 \leq \sigma^2/N$. 

In distributed training setting, $m_{1,t}, m_{2,t}, \cdots, m_{N,t}$ are i.i.d., so $\E [\beta_1 m_{i, t} + (1 - \beta_1) g_{i, t}]$ and $ \E [\sign(\tilde m_{t+1}^i)]$ don't depend on $i$. Assumption \ref{ass::bias_corr} evaluates the discrepancy between the expected value and the expected sign of a measure, positing that the expected values of $\tilde m_t^i$ and $\tilde \sign(m_t^i)$ ought to share the same sign.

%As shown in section~\ref{sec::theory-constrained}, 
We now present our results. 
Similar to the case of global Lion, 
the dynamics of distributed lion can also be divided into two phases 
depending on if the constraint $x \in \mathcal F$ is satisfied. 
% \qq{unify $x \in \mathcal F$ vs. $\norm{\lambda x}_\infty \leq 1$?} \lzchen{unify to $x \in \mathcal F$}
%solves a constrained optimization problem and its dynamics can be divided into 2 phases: 
%\bb\text{Phase I:~~~} \|\lambda x_t\|_\infty > 1,~~~~~~\text{Phase II:~~~} \|\lambda x_t\|_\infty \leq 1. \ee

\paragraph{Phase I ($x \not\in \mathcal F$)} 
In line with the behavior observed in the global Lion model, 
when the constraint is not satisfied, 
both variants of distributed Lion decrease the distance to the feasible set exponentially fast. 
\begin{thm}[Phase I]\label{thm:phsI}
% \qq{what algorithm to follow?}
% \qq{remove $\mathcal K$ completely here.}
% \qq{Define $\mathcal F =\{x \colon \norm{\lambda x} \leq 1 \} $}
Assume % in \eqref{equ:finiteupdate0}
$f\colon \RR^d\to \RR$ is $L$-smooth, $\bone,\btwo \in (0,1)$, and $\btwo>\bone$, and $\epsilon, \lambda > 0$. Let $(x_t)_{t\geq 0}$ be generated by Algorithm~\ref{alg:dist-lion}.
 Define $\mathcal F =\{x \colon \norm{\lambda x}_\infty \leq 1 \} $, and $\dist(x_t, \mathcal{F}) =  \inf_{z \in \mathcal{F}}\norm{z - x_t}$ w.r.t. any norm $\norm{\cdot}$.
%Let $\phi^*$ be the conjugate function of $\phi$. 
For any two non-negative integers $s \leq t$, then $\forall s\leq t$, we have
$$
\dist(x_t, \mathcal{F}) \leq (1-\epsilon\lambda)^{t-s} \dist(x_s, \mathcal{F}).
$$
\end{thm}
Hence, $x_t$ converges to $\mathcal F$ rapidly and stays within $\mathcal F$ once it arrived.

\paragraph{Phase II ($x \in \mathcal F$)}

Now, we present the main result of the analysis for Phase II in Theorems~\ref{thm:majority voting}, \ref{thm:global}, and \ref{thm:averaging}.  
We start with introducing a surrogate metric 
that quantifies the optimality of the solution within Phase II: 
% \qq{I like $\sign$ over $sign$, because we used $\mathrm{dist}$.}
% \qq{$sign$ could mean $s \times i \times g \times n$.} \lzchen{change all $sign$ to $\sign$}
%The proofs are given in appendix~\ref{sec::dist_proof}.
%Define the term \(\Delta_t\) as follows:
\begin{equation}\label{equ::kkt_score}
\mathcal{S}(x) := \langle \nabla f(x), \sign(\nabla f(x)) + \lambda x \rangle.
\end{equation}
% \qq{what is the exact relation with the Lagrangian and KKT condition} 
% \qq{read section 3.1 here \url{https://proceedings.neurips.cc/paper_files/paper/2021/file/f7b027d45fd7484f6d0833823b98907e-Paper.pdf}}
% Given that this analysis pertains to Phase II, we have \(\|\lambda x\|_\infty \leq 1\), and it follows that \(\mathcal{S}(x) \geq 0\).

Let's delve into the implications of \(\mathcal{S}(x) = 0\). 

\begin{pro}
Assume $f$ is continuously differentiable, $\lambda >0$, and $\norm{\lambda x}_\infty \leq 1$. 
Then $\mathcal{S}(x) =0$ implies a KKT stationary condition of $\min_x f(x )~s.t.~\norm{\lambda x}_\infty \leq 1$. 
\end{pro}
% \qq{It should KKT stationary point precisely. Precisely, it means the points that satisfy the first order KKT conditions of problem. Could you explicitly write down the first order KKT conditions and verify exactly?}\lzchen{yes, in appendix} 
This KKT score~\eqref{equ::kkt_score} is tailored to encompass the stationary solutions of the box-constrained problem as described in~\eqref{equ:lionboundc}. Building on this, we then proceed to analyze the convergence for the majority vote, averaging, and global LION strategies throughout this section.
\begin{thm}[Majority Vote] \label{thm:majority voting}
Assumptions \ref{ass::sample}, \ref{ass::L-smooth}, and \ref{ass::bias_corr} hold, consider the Majority vote scheme in Algorithm~\ref{alg:dist-lion}
, $\bone,\btwo \in (0,1)$, and $\btwo>\bone$, and $\sigma \leq 2\sqrt{d}\beta_1 \beta_2^t\|\dd f(x_0)\|, 1\leq t\leq T$ , and $\epsilon, \lambda > 0$. Let $(x_t)_{t\geq0}$ be generated by Majority Vote, and it is in Phase II:  $\|\lambda x_t\|_{\infty} \leq 1$ for all $t$. 

We have 
% \bb
% &\frac{1}{T}\sum_{t=1}^{T} \E \mathcal{S}(x_t) \leq \frac{f(x_0) - f^*}{T\epsilon} + \frac{2\beta_1 \beta_2 \sqrt{d}\|\dd f(x_0)\|}{T(1-\beta_2)}\\  &+ \frac{4\beta_1 L \epsilon d }{1-\beta_2} + \frac{2\sqrt{d}\left((1-\beta_1)\sigma + \sqrt{R^2+C\sigma^2}\right)}{\sqrt{N}}+ 2L\epsilon d,
% \ee
\bb
&\frac{1}{T}\sum_{t=1}^{T} \E\mathcal{S}(x_t) \leq \frac{f(x_0) - f^*}{T\epsilon} + \frac{2D\beta_1 \beta_2 \sqrt{d}\|\dd f(x_0)\|}{T(1-\beta_2)}\\ &+ \frac{4\beta_1 L \epsilon d }{1-\beta_2} +\frac{2\sqrt{d}\sigma(1+\sqrt{C}) + 2\rho}{\sqrt{N}}+ 2L\epsilon d,
\ee
where $C = \beta_1^2 (1-\beta_2)\frac{1}{1+\beta_2} + (1-\beta_1)^2$, $D = \max\{1, \sigma/\left(2\sqrt{d}\beta_1 \beta_2^T\|\dd f(x_0)\|\right)\}$,
\bb
\rho_t[k] =
    \begin{cases}
      0 ~~~~~~~~~~~~~~~~~~~~~~~~\text{if~ $\E [\sign(\tilde m_{t+1}^i[k])]=0$,} \\
      \E [\tilde m_{t+1}^i[k]] / \E [\sign(\tilde m_{t+1}^i[k])] ~~~~~~~~~~\text{else,}
    \end{cases}  
\ee, and $\rho = \max_{1\leq t \leq T}\norm{\rho_t}$.
\end{thm}
The result above shows that $\frac{1}{T}\sum_{t=1}^{T} \E \mathcal{S}(x_t)$ decays with an  $\mathcal{O}(\frac{1}{T\epsilon}+\frac{1}{T(1-\beta_2)} + \epsilon + \frac{1}{\sqrt{N}})$.
This rate is in fact on par with global Lion as we show in the following result: 
%as we show in Theorem ~\ref{thm:global}. %\red{reference?}.  
%For Global Lion, the convergence analysis in Phase II is shown as the following theorem: % \ref{thm:global}:
\begin{thm}[Global] \label{thm:global}
Assumptions \ref{ass::sample} and \ref{ass::L-smooth} hold, Consider the scheme in Algorithm~\eqref{equ:global}, with the same settings in Theorem~\ref{thm:majority voting},
we have
\bb
\frac{1}{T}\sum_{t=1}^{T} \E \mathcal{S}(x_t) &\leq \frac{f(x_0) - f^*}{T\epsilon} + \frac{2\beta_1 \beta_2 \sqrt{d}\|\dd f(x_0)\|}{T(1-\beta_2)}\\
&+ \frac{4\beta_1 L \epsilon d }{1-\beta_2} + \frac{2(1-\beta_1)\sqrt{d} \sigma}{\sqrt{N}} + 2L\epsilon d. 
\ee
\end{thm} 
%On the other hand, we have a looser bound on the distributed lion with averaging rule, 
%D-Lion (Avg), 
\begin{thm}[Averaging] \label{thm:averaging}
Assumptions \ref{ass::sample} and \ref{ass::L-smooth} hold, consider the Averaging scheme in Algorithm~\ref{alg:dist-lion}  
, with the same settings in Theorem~\ref{thm:majority voting}, 
%\qq{assumptions} 
we have
\bb
\frac{1}{T}\sum_{t=1}^{T} &\E\mathcal{S}(x_t) \leq \frac{f(x_0) - f^*}{T\epsilon} + \frac{2\beta_1 \beta_2 \sqrt{d}\|\dd f(x_0)\|}{T(1-\beta_2)}\\
&+ \frac{4\beta_1 L \epsilon d }{1-\beta_2} + \frac{2\beta_1 \sqrt{d}\sigma}{\sqrt{1+\beta_2}} + 2(1-\beta_1)\sqrt{d}\sigma+ 2L\epsilon d
\ee
\end{thm}
The Averaging method's convergence bound doesn't improve with more workers since \(\frac{1}{N}\sum_{i=1}^N \sign(\delta_{i,t})\) doesn't approximate \(\sign(\sum_{i=1}^N \delta_{i,t})\) effectively, unlike the Majority Vote's approach $\sign(\sum_{i=1}^N \sign(\delta_{i,t}))$.

\section{Experiment}
\label{sec::exp}
In this section, we perform a thorough evaluation of the \mavolion{} algorithm, employing both the averaging and majority vote aggregation methods. The design of our experiments is aimed at addressing the following questions to ascertain the algorithm's efficacy and performance:
\begin{itemize}
  \itemsep0em  % This sets the space between items to 0em. Adjust the value as needed.

\item \textbf{(Q1)} How does \mavolion{} stand in comparison to global distributed training approaches, i.e., methods that aggregate gradients from local workers and employ an optimizer on the collective gradient?

\item \textbf{(Q2)} How does \mavolion{} perform when compared to established communication-efficient distributed training methodologies?

\item \textbf{(Q3)} How does \mavolion{} scale on large vision or language problems?
\end{itemize}

\begin{figure*}[t!]
    \centering
    \includegraphics[width=\textwidth]{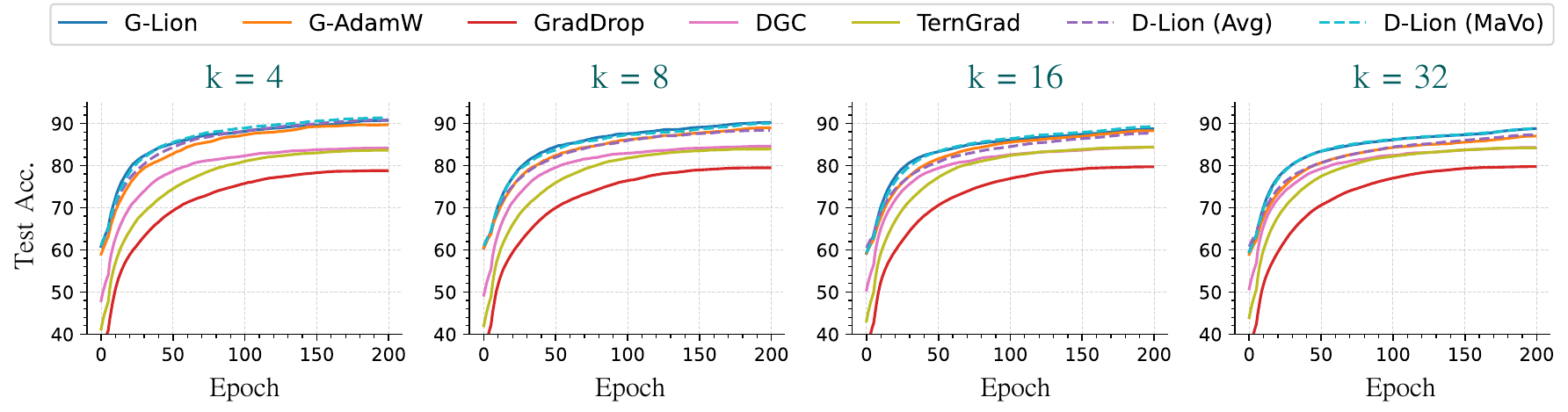}
    \caption{Performance of \mavolion{} v.s. other efficient distributed optimizers on CIFAR-10 with 4, 8, 16, and 32 workers, each worker at each iteration runs on a local batch with size 32. All results are averaged over three seeds.
    }

    \label{fig:cifar}
\end{figure*}

\subsection{Comparing \mavolion{} Against Established Methods on CIFAR-10}

To address \textbf{Q1} and \textbf{Q2}, we compare \mavolion{} with both the averaging and the majority vote methods, against established low-bandwidth distributed training techniques and the global distributed training methods. We consider the following baseline methods: \textbf{1) Global \adamw{} (G-\adamw{})}, where we apply \adamw{} with the averaged gradients from all workers. \textbf{2) Global \lion{} (G-\lion{})}, where we apply \lion{} with the averaged gradients from all workers. Note that Global \adamw{} and Global \lion{} serve as the performance and communication upper bounds. \textbf{3) \mavolion{} with Averaged Updates (D-\lion{} (Avg))},  In contrast to the majority vote mechanism used in \mavolion{}, this variant averages the binary update vectors from all workers. While D-\lion{} (Avg) might offer improved performance in principle, it comes at the cost of non-binary communication from the server to the workers. \textbf{4) TernGrad}~\citep{wen2017terngrad}. The main idea is to tenarize the gradient into a vector of $\{-1, 0, 1\}$, which is similar to what \lion{} does. But this process is done on the gradient level instead of on the update level \textbf{5) Gradient Dropping (GradDrop)}~\citep{aji2017sparse}. The main idea is to drop insignificant gradient entries and only transmit sparse gradient signals. \textbf{6) Deep Gradient Compression (DGC)}~\citep{lin2017deep}. DGC is built on top of the GradDrop, but additionally applies momentum correction, local gradient clipping, momentum factor masking, and warm-up training. 

\paragraph{Experiment Setup} For GradDrop, DGC, and TernGrad, we choose the compression rate of $0.04$ (note that $1/32 = 0.03125$) to match the bandwidth of the D-\lion{} (MaVo). We conduct experiments on the CIFAR-10 dataset using a vision transformer (ViT) with 6 layers, 8 heads, and a hidden dimension of 512. This is because ViT has arguably become the most widely used architecture in computer vision, and we empirically found no additional gain in performance when using a larger ViT on CIFAR-10. In addition, to validate how \mavolion{} performs with different numbers of workers, we consider $k \in \{4, 8, 16, 32\}$, each worker at each iteration will sample an i.i.d data batch of size 32.  

We list the optimal hyperparameters selected for each method from Figure~\ref{fig:cifar} in Table~\ref{tab:hyperparameters}. The learning rates are selected from $\{0.00005, 0.001, 0.005, 0.01\}$ and the weight decays are selected from $\{0.0005, 0.001, 0.005\}$. For each experiment, we use a cosine learning rate scheduler and run for 200 epochs, and we ensure that in each epoch, each local worker sees the entire dataset once.

\begin{table}[h!]
\centering
 \resizebox{0.95\linewidth}{!}{
\begin{tabular}{lrrc}
\toprule
Method &  lr $\epsilon$ & wd $\lambda$ & compression rate \\
\midrule
G-AdamW & 0.0001  & 0.0005 &  - \\
G-Lion & 0.00005 & 0.005 &  - \\
DGC    & 0.01    & 0.0005 &  0.96 \\
GradDrop & 0.001  & 0.0005 &  0.96 \\
TernGrad & 0.001  & 0.0005 &  - \\
D-Lion (Avg) & 0.00005  & 0.005 &  - \\
D-Lion (MaVo) & 0.00005  & 0.005 &  - \\
\bottomrule
\end{tabular}
}
\caption{\textbf{Hyperparameters} for each method in Figure~\ref{fig:cifar}. Where lr represents learning rate and wd represents weight decay.}
\label{tab:hyperparameters}
\end{table}

Each experiments are conducted with three random seeds $\{42, 52, 62\}$, which results in a total of $4 \times 7 \times 3 = 84$ experiments.

\begin{figure}[h!]
    \centering
    \includegraphics[width=\columnwidth]{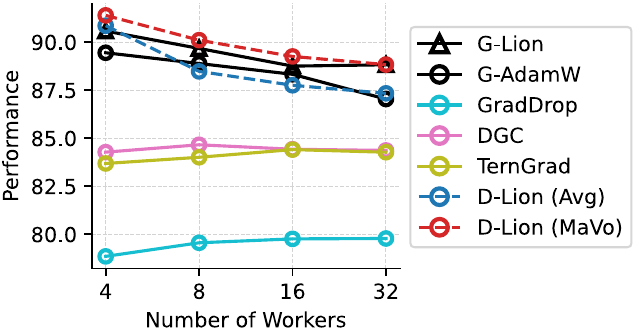}
   
    \caption{Performance of different methods v.s. $k$.}
    \label{fig:workers}
    
\end{figure}
\begin{figure}[h!]
    \centering
    \includegraphics[width=1.0\columnwidth]{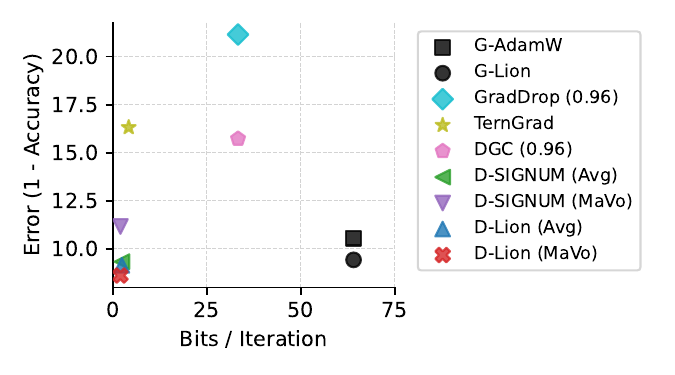}
   
    \caption{Test Error v.s. Communication Bits per Iteration (closer to the lower-left is better). Note that we set G-Lion and G-AdamW are both 64, because they require 32 bits per parameter, and there are both worker-to-server and server-to-worker communications. }
    \label{fig:cifar-bandwidth}
    
\end{figure}

\myobs{Observation} We plot the testing accuracy (Test Acc.) over epochs for different methods in Figure~\ref{fig:cifar}, the best testing accuracy of different methods over the number of workers in Figure~\ref{fig:workers}, and the performance versus per-iteration bandwidth in Figure~\ref{fig:cifar-bandwidth} when using $k=4$ workers. From the above plots, we make the following observations.
\begin{itemize}
  \itemsep0em  % This sets the space between items to 0em. Adjust the value as needed.
    \item Compared to global distributed training methods, D-Lion (MaVo) performs on par with G-Lion. D-Lion (Avg) performs slightly worse than G-Lion but is on par with G-Adamw (Figure~\ref{fig:cifar}).
    \item Compared to established communication efficient distributed training methods, both D-Lion (MaVo) and D-Lion (Avg) outperform GradDrop, DGC and TernGrad by a large margin (Figure~\ref{fig:cifar}).
    \item We observe that both D-Lion (MaVo) and D-Lion (Avg) exhibit strong performance while being 30x more communication efficient than global distributed training methods like G-AdamW. To broaden our comparison, we introduced two additional baseline methods: \textbf{D-SIGNUM (Avg)} and \textbf{D-SIGNUM (MaVo)}. These baselines apply our proposed techniques to the SIGNUM framework instead of Lion.\footnote{Note that D-SIGNUM (Avg/MaVo) further subsumes D-SignSGD~\citep{bernstein2018signsg1,bernstein2018signsgd2}.} We set $\beta=0.99$ for D-SIGNUM. According to our results, depicted in Figure~\ref{fig:cifar-bandwidth}, these SIGNUM-based methods do not perform as well as their Lion-based counterparts.
    \item We notice that the overall performance of the same optimizer becomes worse as $k$ goes larger, this is consistent with the observation made in DGC~\citep{lin2017deep}. We hypothesize that this may be due to the larger effective batch size resulting in smaller stochasticity, which is consistent with why D-Lion (MaVo) performs a bit better than G-\lion{} on CIFAR-10 (Figure~\ref{fig:workers}).
\end{itemize}

% \begin{table*}[!hpbt]
% \begin{center}
% % \resizebox{\linewidth}{!}{
%  \setlength{\tabcolsep}{6.5mm}
% \begin{tabular}{l|c|ccc}
% \Xhline{3\arrayrulewidth} 
% {Model}  &   AdamW   & G-Lion   & D-Lion(MaVo) &    D-Lion(Avg)     \\ \hline 
%   ViT-S/16 &  79.74   &  \underline{79.82}  &  79.69\cellcolor{Graylight}  & \cellcolor{Graylight} \textbf{80.11}  \\
%     ViT-B/16 &  80.94   &  \underline{80.99}  &  80.79\cellcolor{Graylight} & \cellcolor{Graylight} \textbf{81.13}  \\
%     GPT-2++ (350M)  &  2.914   &  \textbf{2.910}  &  \underline{2.911}\cellcolor{Graylight} & \cellcolor{Graylight} 2.912  \\
%     GPT-2++ (760M)  &  2.688   &  \textbf{2.685}  &  \textbf{2.685}\cellcolor{Graylight} & \cellcolor{Graylight} 2.687  \\
% \Xhline{3\arrayrulewidth} 
% \end{tabular}
% % }
% \caption{Main results on various of large-scale benchmarks. All the experiments are conduct on 32 workers. For ViT experiments, we evaluate the Top-1 accuracy and for GPT2++ experiments we show the validation loss. We mark the best performance with bold text and the second one with an underline. }
% \label{tab:result}
% % \vspace{-5pt}
% \end{center}
% \end{table*}

\begin{table*}[!hpbt]
\begin{center}
\setlength{\tabcolsep}{4mm} % Adjusted for better spacing

\begin{tabular}{lcccc}
\toprule
\multirow{2}{*}{Method} & \multicolumn{2}{c}{Image Classification} & \multicolumn{2}{c}{Language Modeling} \\ 
\cmidrule(lr){2-3} \cmidrule(lr){4-5}
 & ViT-S/16 & ViT-B/16 & GPT-2++ (350M) & GPT-2++ (760M) \\ 
\midrule 

AdamW & 79.74 & 80.94 & 18.43 & 14.70 \\

G-Lion & \underline{79.82} & \underline{80.99} & \textbf{18.35} & \textbf{14.66} \\

D-Lion (MaVo) & 79.69\cellcolor{Graylight} & 80.79\cellcolor{Graylight} & \underline{18.37}\cellcolor{Graylight} & \textbf{14.66}\cellcolor{Graylight} \\

D-Lion (Avg) & \textbf{80.11}\cellcolor{Graylight} & \textbf{81.13}\cellcolor{Graylight} & 18.39\cellcolor{Graylight} & 14.69\cellcolor{Graylight} \\
\bottomrule
\end{tabular}
\caption{Results on ImageNet classification and OpenWebText language modeling. For ImageNet experiments, we report the Top-1 accuracy. For language modeling experiments, we report the validation perplexity. The best performance is marked with bold text, and the second best with an underline.}
\label{tab:result}

\end{center}
\vspace{-5pt}
\end{table*}

\begin{table*}[!hpbt]
\begin{center}
% \resizebox{\linewidth}{!}{
 \setlength{\tabcolsep}{3.5mm}
\begin{tabular}{lccccccc}
% \Xhline{3\arrayrulewidth} 
\toprule
Method  &   Arc-Easy   & Arc-Challenge   & BoolQ &  PIQA & SIQA & HellaSwag & OBQA      \\ 
\midrule 
   0-Shot &  76.64  & 43.06 &  76.43   & 78.64 & 45.96 & 56.87 & 33.53 \\
   \midrule
 G-AdamW  &  \underline{77.06}  & \textbf{46.06} &  \underline{77.23}   & \textbf{79.18} & 48.97 & \textbf{59.23} & \underline{35.51} \\
G-Lion  &  \textbf{77.11}  & 45.54 &  \textbf{77.50}   & \textbf{79.18} & \underline{49.64} & 58.93 & \underline{35.51} \\
\cellcolor{Graylight}D-Lion (MaVo) &  \cellcolor{Graylight} 76.86  & \cellcolor{Graylight} \underline{45.72} & \cellcolor{Graylight} 77.14   & \cellcolor{Graylight} 78.92 & \cellcolor{Graylight} \textbf{49.75} &\cellcolor{Graylight} \underline{58.96} & \cellcolor{Graylight} \textbf{35.71} \\
\cellcolor{Graylight}D-Lion (Avg) & \cellcolor{Graylight} 76.35  & \cellcolor{Graylight} 45.54 & \cellcolor{Graylight}  76.90  & \cellcolor{Graylight} 78.76 &  \cellcolor{Graylight} 48.06 &\cellcolor{Graylight} 59.06  &\cellcolor{Graylight} 32.14 \\
\bottomrule
\end{tabular}
% }
\caption{3-Shot instruction finetuning downstream evaluation results on various datasets. We mark the best performance with bold text and the second one with an underline.}
\vspace{-5pt}
\label{tab:result2}

\end{center}
\end{table*}

\subsection{Scale to Larger Models on Larger Datasets}
To answer \textbf{Q3}, we validate \mavolion{} on several large-scale setups including both vision and natural language processing tasks. Under this setting, we compare D-Lion (MaVo) and D-Lion (Avg) against G-AdamW and G-Lion. For the vision task, we tested ViT-S/16~\citep{dosovitskiy2020image} and ViT-B/16 on the ImageNet-1K~\citep{imagenet15russakovsky} classification benchmark. For the natural language processing task, we perform both language pretraining and finetuning tasks. This is because Lion has shown good results on language modeling. For the language model pretraining task, we pretrain GPT2++~\citep{radford2019language} (the GPT-2 model with modern training techniques adopted from the LLaMA model~\citep{touvron2023llama}) on the OpenWebText~\citep{Gokaslan2019OpenWeb} benchmark, for both 350M and 760M size models. For the language model finetuning task, we conduct few-shot finetuning of the LLaMA 7B model~\citep{touvron2023llama} and evaluate the models' downstream performance on standard downstream evaluation benchmarks~\citep{Clark2018ThinkYH, zellers2019hellaswag, clark2019boolq, Mihaylov2018CanAS, bisk2020piqa,sap2019socialiqa}.

\paragraph{Experiment Setup} 
For the ImageNet-1K benchmark, we train all methods for 300 epochs, using a global batch size of 4096 and data augmentations MixUp~\cite{zhang2017mixup} of 0.5 and AutoAug~\cite{cubuk2018autoaugment}. When training ViT-S/16, we use a learning rate of $3e^{-3}$ for G-AdamW, with betas of $(0.9, 0.999)$ and a weight decay of 0.1. For G-Lion, D-Lion (MaVo), and D-Lion (Avg), we use a learning rate of $3e^{-4}$, betas of $(0.9, 0.99)$, and a weight decay of 1.0. As for ViT-B/16, we use a learning rate of $1e^{-3}$ for G-AdamW, with betas of $(0.9, 0.999)$ and a weight decay of 1.0, while for all Lion variants, we use a learning rate of $1e^{-4}$, betas of $(0.9, 0.99)$, and a weight decay of 10.0.
For pretraining language models on the OpenWebText dataset, we build GPT2++ models using the original GPT2 model, but with modern training techniques from the LLaMA model, including using the Gated Linear Unit activation for the multilayer layer perceptron layers (MLPs) and the RMSNorm~\citep{zhang2019root} instead of the LayerNorm~\citep{ba2016layer}. Following the Chinchilla scaling law~\citep{hoffmann2022training}, we trained the 350M model for 14,000 iterations and the 760M model for 30,000 iterations, both with 1,024 tokens. For G-AdamW, we use a learning rate of $3e^{-4}$, betas of $(0.95, 0.99)$, and a weight decay of 0.1. For all Lion variants, we use a learning rate of $9e^{-5}$, betas of $(0.9, 0.99)$, and a weight decay of 1.0. All the models are trained under a global batch size of 480. For the instruction finetuning task, we instruct finetune a LLaMA 7B model for 3 epochs with batch size 32. We use $2e^{-5}$ learning rate, betas of $(0.9, 0.999)$, 0 weight decay for G-AdamW and $6e^{-6}$, $(0.9, 0.99)$ betas, $0.01$ weight decay for all Lion variants. For all pretraining experiments, we use $4 \text{nodes} \times 8 \text{gpus} = 32$ workers. For instruction finetuning experiments, we use 4 workers per experiment. 

\myobs{Observation}  We summarize the results in Table~\ref{tab:result} (ImageNet 1K and OpenWebText Language Model Pretraining) and Table~\ref{tab:result2} (Instruction Finetuning). From these two tables, it is evident that both D-Lion (Avg) and D-Lion (MaVo) can maintain a performance similar to, or even better than, that of G-AdamW and G-Lion, on both large-scale vision and language tasks. We observe that D-Lion (Avg) outperforms D-Lion (MaVo) on ImageNet, and observe the opposite on language modeling and instruction finetuning. We hypothesize that these differences are due to the impact of global batch size. As a result, we recommend using D-Lion (Avg)~/~(MaVo) when the global batch size is large~/~small.%\red{check}
\section{Conclusion and Future Work} \label{sec::conclusion}
In this paper, we introduced \mavolion{}, a communication-efficient distributed training strategy that builds upon the \lion{} optimizer's binary update mechanism. \mavolion{} is designed to minimize communication overhead by allowing workers to independently manage their optimizer states and exchange only binary or low-precision update vectors with the server. We proposed two aggregation techniques within the \mavolion{} framework: average-based (\mavolion{} Avg) and majority vote-based (\mavolion{} MaVo) algorithms. We provide both theoretical and empirical results to demonstrate \mavolion{}'s effectiveness, scalability, and efficiency. Notably, we show that \mavolion{} performs significantly better than existing communication-friendly methods. In the meantime, \mavolion{} demonstrates performance on par with strong global distributed training baselines, while being 32x more communication efficient. As our method is orthogonal to existing communication-efficient methods, an interesting future direction is to combine both techniques from both worlds for further improvement. As a limitation, currently \mavolion{} (Avg~/~MaVo) performs inconsistently across different datasets and benchmarks, it will be an interesting future research direction to understand when and why one performs better than the other.

\section{Broader Impact}
This paper presents a novel method that aims to improve distributed training. While we acknowledge that our work could have a multitude of potential societal consequences, we do not believe any specific ones need to be highlighted.
\bibliography{example_paper}
\bibliographystyle{icml2023}
%%%%%%%%%%%%%%%%%%%%%%%%%%%%%%%%%%%%%%%%%%%%%%%%%%%%%%%%%%%%%%%%%%%%%%%%%%%%%%%
%%%%%%%%%%%%%%%%%%%%%%%%%%%%%%%%%%%%%%%%%%%%%%%%%%%%%%%%%%%%%%%%%%%%%%%%%%%%%%%
% APPENDIX
%%%%%%%%%%%%%%%%%%%%%%%%%%%%%%%%%%%%%%%%%%%%%%%%%%%%%%%%%%%%%%%%%%%%%%%%%%%%%%%
%%%%%%%%%%%%%%%%%%%%%%%%%%%%%%%%%%%%%%%%%%%%%%%%%%%%%%%%%%%%%%%%%%%%%%%%%%%%%%%
\newpage
\appendix
\onecolumn
\section{Appendix I}\label{pdx}
% \subsection{Additional Experiment Details}
% \label{sec::apx-exp}
% In this section, we provide additional experiment details.
% \paragraph{CIFAR Experiments}
% We list the optimal hyperparameters selected for each method from Figure~\ref{fig:cifar} in Table~\ref{tab:hyperparameters}. The learning rates are selected from $\{0.00005, 0.001, 0.005, 0.01\}$ and the weight decays are selected from $\{0.0005, 0.001, 0.005\}$. For each experiment, we use a cosine learning rate scheduler and run for 200 epochs, and we ensure that in each epoch, each local worker sees the entire dataset once.

% \begin{table}[h!]
% \centering
% \begin{tabular}{@{}l|ccc@{}}
% \toprule
% Method &  Learning Rate $\epsilon$ & Weight Decay $\lambda$ & Compression Rate \\
% \midrule
% G-AdamW & 0.0001  & 0.0005 &  - \\
% G-Lion~\citep{chen2023lion}  & 0.00005 & 0.005 &  - \\
% DGC~\citep{lin2017deep}     & 0.01    & 0.0005 &  0.96 \\
% GradDrop~\citep{aji2017sparse} & 0.001  & 0.0005 &  0.96 \\
% TernGrad~\citep{wen2017terngrad} & 0.001  & 0.0005 &  - \\
% D-Lion (Avg) & 0.00005  & 0.005 &  - \\
% D-Lion (MaVo) & 0.00005  & 0.005 &  - \\
% \bottomrule
% \end{tabular}
% \caption{\textbf{Hyperparameters} for each method in Figure~\ref{fig:cifar}.}
% \label{tab:hyperparameters}
% \end{table}

This section is focusing on the proof of \lion{} dynamics, and will be organized into these folders: 
\begin{itemize}
\item Phase I: 
\begin{itemize}
    \item Constraint enforcing: Discrete time
\end{itemize}
\item Phase II: 
\begin{itemize}
    \item Majority Voting convergence
    \item Avg update convergence
    \item Global LION convergence
\end{itemize}
\end{itemize}
% Given a $d$-dimensional vector $v$ that satisfies $\|v\|\leq R$, we denote a stochastic sign operator as~\cite{sun2023momentum}
% \bb
% [\mathcal{S}_R(v)]_i
% = \begin{cases}
% -sign(v_i) & \text{with $p = \frac{1}{2} - \frac{|v_i|}{2R}$} \\  
% sign(v_i) & \text{with $p = \frac{1}{2} + \frac{|v_i|}{2R}$}  .
% \end{cases}
% \ee

% In distributed training, the learning objective we aim to minimize:
% \begin{equation}
% \label{equ:distibuted-objective}
% \min_x f(x) = \frac{1}{N} \sum_{i=1}^N \mathbb{E}_{\xi_i \sim \mathcal{D}_i} \bigg[ f(x; \xi_i)\bigg].
% \end{equation}
% \qq{the decision to repeat Eq (2), but not the algorithm 1 here seems arbitrary. what are do you intend to do with Eq (10) here?}\lzchen{recap what we are solving, it's }
% %
In line with the behavior observed in the global \lion{} approach, \lion{} under a distributed setting also exhibits the two phases. In Section~\ref{pix_sec::p1}, we show that converging to box can be exponentially fast using our Algorithm~\ref{alg:dist-lion}. We start with introducing a notion of KKT score function that quantifies a stationary solution to the box constrained optimization problem \eqref{equ:lionboundc} in Section~\ref{pix_sec::p2}. 
Building on this, we then proceed to analyze the convergence in terms of the KKT score function for the majority vote  (Section \ref{sec::maj_v}), averaging  (Section \ref{sec::avg}), and global LION strategies  (Section \ref{sec::glob}). 
%throughout. 

\subsection{Phase I: Constraint Enforcing}\label{pix_sec::p1}

%\qq{Say something here. This is meant to be read by human, who needs purpose. purpose, purpose, purpose.}\lzchen{added purpose}

%\qq{We will study phase i, we show that ... }\lzchen{added purpose}
We  study phase I in this section. We  show that when the constraint is not satisfied, 
both variants of distributed Lion decrease the distance to the feasible set exponentially fast. 

%\qq{Algorithm 1, not algorithm 1. correct accordingly}\lzchen{changed all to Algorithm 1}
\begin{thm}[Phase I]\label{pix_thm:phsI}

Assume 
$f\colon \RR^d\to \RR$ is $L$-smooth, $\bone,\btwo \in (0,1)$, and $\btwo>\bone$, and $\epsilon, \lambda > 0$, and $1-\epsilon \lambda \in (0,1)$. Let $(x_t)_{t\geq 0}$ be generated by Algorithm~\ref{alg:dist-lion}. Define $\mathcal F =\{x \colon \norm{\lambda x}_\infty \leq 1 \} $, and $\dist(x_t, \mathcal{F}) =  \inf_{z \in \mathcal{F}}\norm{z - x_t}$ w.r.t. any norm $\norm{\cdot}$.

For any two non-negative integers $s \leq t$, then $\forall s\leq t$,  we have
$$
\dist(x_t, \mathcal{F}) \leq (1-\epsilon \lambda)^{t-s} \dist(x_s, \mathcal{F}).
$$

\end{thm}
\begin{proof}
Recall Algorithm \ref{alg:dist-lion}:
\bb
\delta_{i,t} &\leftarrow \sign\big(\bone m_{i,t} + (1 - \bone)\nabla_x f(x_{t};
\xi_{i,t})\big)\\
m_{i,t+1} &\leftarrow \btwo m_{i,t} + (1 - \btwo) \nabla_x f(x_{t}; \xi_{i,t})\\
\Delta_t &= \begin{cases}
    \frac{1}{N} \big(\sum_{i=1}^N \delta_{i,t}\big) & \text{(Averaging)} \\
    \sign \big(\sum_{i=1}^N \delta_{i,t}\big) & \text{(Majority Vote)} \\
\end{cases}\\
x_{t+1} &= x_t - \epsilon (\Delta_t + \lambda x_t)
\ee

Rewrite the update into the following form: 
$$
x_{t+1} = (1- \epsilon \lambda) x_t -  \epsilon \Delta_t,
$$

Define  $w_{s\to t} = (1-\epsilon \lambda)^{t-s}$. 
Unrolling this update  yields,
\bb 
x_{t}
& =(1- w_{s\to t}) z_{s\to t}
+ w_{s\to t} x_s, ~~~~~~ 
~~~~ 
z_{s\to t} = \frac{\sum_{k=s}^{t-1} w_{k\to t} (-\Delta_t/\lambda)}{\sum_{k=s}^{t-1} w_{k\to t}}. 
\ee

We have $z_{s\to t} \in \mathcal{F}$ since $-\Delta_t/\lambda \in \mathcal{F}$. For any $\epsilon>0$, let $\hat x_s \in \mathcal{F}$ be the point satisfying $\norm{\hat x_s - x_s} \leq \dist(x_s, \mathcal{F}) + \eta$.
Hence, we have
\bb 
\mathrm{dist}(x_t, ~ \mathcal{F}) 
& = \inf_{z \in \mathcal{F}} \norm{x_t  - z }  \\
& \leq \norm{ x_t - (1-w_{s\to t}) z_{s\to t} - w_{s\to t} \hat x_s)} \\
& =  w_{s\to t} \norm{x_s -\hat x_s} \\
& \leq (1-\epsilon \lambda)^{t-s}  (\dist(x_s, \mathcal{F}) + \eta).
\ee 
As $\eta\to 0$, we achieve the desired result.
\end{proof}
\subsection{Phase II}\label{pix_sec::p2}
We study the convergence of Phase II in this section. We begin by defining a KKT score function to quantify stationary solutions for the box-constrained optimization problem discussed in Section~\ref{pix_sec::p2}. Following this, we analyze convergence through the KKT score across majority vote (Section~\ref{sec::maj_v}), averaging (Section~\ref{sec::avg}), and global Lion strategies (Section~\ref{sec::glob}).

First, we list the following assumptions used in our proof. 

\begin{ass}[Smooth and Differentiable $f$]\label{pdx_ass::L-smooth}
    Function $f(\cdot)$ is differentiable and L-smooth.
\end{ass}

\begin{ass}[Variance bound]\label{pix_ass::variance} $\mathcal{D}_i$ is i.i.d. drawn from a common distribtion $\mathcal \pi_*$, and the stochastic sample $\xi^i \sim \mathcal{D}_i$ is i.i.d. and upon receiving query $x \in \mathbb{R}^d$, the stochastic gradient oracle gives us an \emph{independent} unbiased estimate $\dd f(x;\xi^i)$ from the  $i$-th worker that has coordinate bounded variance:
	\begin{equation*}
	\mathbb{E}_{\xi}[\dd f(x;\xi^i)]= \dd f(x), \qquad   \E_\xi\left[\|\dd f(x;\xi^i)-\dd f(x)\|^2\right]  \leq \sigma^2.
	\end{equation*}
\end{ass}

\begin{ass}[Bias Correction]\label{pix_ass::bias_corr}
  Consider the sequence $\{m_t^i\}_{t>0,i \in [N]}$ generated by Algorithm \ref{alg:dist-lion}, $\E [\tilde m_{t}^i] / \E [\sign(\tilde m_{t}^i)] \geq 0$.
\end{ass}
Here we define the a KKT score function for box constrained problem \eqref{equ:lionboundc}:
\bb
\mathcal{S}(x) := \langle \nabla f(x), \sign(\nabla f(x)) + \lambda x \rangle.
\ee
\begin{pro}

Assume $f$ is continuously differentiable, $\lambda >0$, and $\norm{\lambda x}_\infty \leq 1$. 
Then $\mathcal{S}(x) =0$ implies a KKT stationary condition of $\min_x f(x )~s.t.~\norm{\lambda x}_\infty \leq 1$. 
\end{pro}
\begin{proof}

We will verify that $\mathcal{S}(x)=0$ coincides with the first order KKT conditions of the box constrained optimization problem \eqref{equ:lionboundc}.

Recall the box constrained problem in \eqref{equ:lionboundc}, we can rewrite it into the following formulation:
\bb
\min_{x\in\RR^d} f(x)~~~~s.t.~~~~ \lambda x_i -1 \leq 0,~~~ -\lambda x_i -1 \leq 0,~~~ \forall ~i \in [d].
\ee 
Let $\mu = (\mu_1, \mu_2, \cdots, \mu_d)\tt$ and $\tilde\mu = (\tilde\mu_1, \tilde\mu_2, \cdots, \tilde\mu_d)\tt$, then its first order KKT stationary condition can be written as:
\bb
&\partial_{x_i} f(x) + \mu_i \lambda - \tilde \mu_i \lambda =0&\ant{Stationarity}\\
&\mu_i (\lambda x_i -1) = 0, ~~~\tilde \mu_i (-\lambda x_i -1) =0&\ant{Complementary slackness}\\
&\mu_i \geq 0, ~~~\tilde \mu_i \geq 0&\ant{Dual feasibility}\\
&\lambda x_i -1 \leq 0,~~~ -\lambda x_i -1 \leq 0 &\ant{Primal feasibility}\\
& \forall ~i \in \{1,2,\cdots, d\}.
\ee
Expressing \(\mathcal{S}(x)\) element-wisely, we obtain:
\begin{align*}
\mathcal{S}(x) = \sum_{k=1}^d  \mathcal{S}_k(x), &&\text{with}&&
\mathcal{S}_k(x) = \partial_{x_k} f(x) \cdot \left(\sign(\partial_{x_k} f(x)) + \lambda x_k\right), 
\end{align*} 
where $x_k$ denotes the $k$-th element of vector $x$.  
Since $\norm{\lambda x}_\infty \leq 1$, we have $\mathcal{S}_k(x) \geq 0$, because 
\bb
\mathcal{S}_k(x) &= \partial_{x_k} f(x) \cdot \left(\sign(\partial_{x_k} f(x)) + \lambda x_k\right) \\
               &= |\partial_{x_k} f(x)| + \lambda \partial_{x_k} f(x) \cdot x_k \\
               &\geq |\partial_{x_k} f(x)| - |\partial_{x_k} f(x)| \cdot |\lambda x_k|\\
               &=|\partial_{x_k} f(x)| (1 - |\lambda x_k|) \\
               &\geq 0 \ant{since $\norm{\lambda x}_\infty \leq 1.$}
\ee

Hence, if \(\mathcal{S}(x)= 0\), we have $\mathcal{S}_k(x) =0$ 
for each component \(k\). It means that we have either \(\sign(\partial_{x_k} f(x)) + \lambda x_k = 0\) or \(\partial_{x_k} f(x) = 0\) for each coordinate $k$.

There are two primary cases to consider for each $k$: 

\begin{itemize}
\item \textbf{Case I}: \(\partial_{x_k} f(x) = 0\). This suggests that we reach a stationary condition of $f(x)$ w.r.t. coordinate $x_k$,  and the KKT condition is satisfied in this case with  $\mu_k = \tilde \mu_k = 0$.

\item \textbf{Case II}: \(\sign(\partial_{x_k} f(x)) + \lambda x_k = 0\), it follows that \(x_k = -\frac{1}{\lambda} \sign(\partial_{x_k} f(x))\). 
\begin{itemize}
    \item if $\sign(\partial_{x_k}f(x) = 1$, then $\partial_{x_k}f(x) \geq 0$, and the KKT condition is satisfied with $\mu_k = 0$ and $\tilde \mu_k = \partial_{x_k} f(x) / \lambda$
    \item if $\sign(\partial_{x_k}f(x)) = -1$, then $\partial_{x_k}f(x) \leq 0$, and the KKT condition is satisfied with $\tilde\mu_k = 0$ and $\mu_k = \partial_{x_k} f(x) / \lambda$. 
\end{itemize}
\end{itemize}

It turns out the two cases above exactly covers the KKT stationary solution pair $(x, \mu, \tilde \mu)$ of the box constrained problem in \eqref{equ:lionboundc}. 

In conclusion, \(\mathcal{S}(x) = 0\) signifies reaching a stationary point of the bound-constrained optimization problem, as formulated in \eqref{equ:lionboundc},  providing critical insights into the convergence behavior of the algorithm under consideration.
\end{proof} 
\subsubsection{Majority Vote}\label{sec::maj_v}
%\qq{the algorithm scheme should appear early. before the phase I analysis at least}
%\qq{text need to be well organized.}\lzchen{I removed all the implicit forms, only left the explicit update here now.}
Assume $f\colon \RR^d\to \RR$ is $L$-smooth, and $N$ is the number of workers, on the $i$-th worker, 
%\qq{Assume that xx, and xx, and xxx.}\lzchen{fixed}
consider the following  scheme based on the majority vote: 
%\qq{define the scheme in the very begining of the appendix (before Theorem A.1)?}\lzchen{It might be better to define the scheme at the beginnings of each subsection, in the following sections, we firs define a scheme and the give its convergence analysis, meanwhile, Theorem A.1 doesn't depends on update scheme}
%\qq{use explicit form?}\lzchen{unified to explicit form.}
\bbb \label{equ:alldisc} 
\begin{split}
& g_t^i : = \dd f(x_t;\xi_t^i)\\
& m_{t+1}^i = \beta_2m_t^i + (1-\beta_2) g^i_t\\
& \tm_{t+1}^i = \beta_1 m_t^i + (1-\beta_1) g^i_t  \\
& x_{t+1} = x_t - \epsilon \left(\sign\left(\sum_{i=1}^N \sign(\tilde m^i_{t+1})\right) + \lambda x_{t} \right).  \ant{Majority Voting} %\qq{\text{small bracket outside of large one, arguly}} \lzchen{\text{sorry, i corrected it}} \\
\end{split} 
\eee

\begin{thm}[Convergence in Phase II] \label{pix_thm:majority voting}
Assumption \ref{pdx_ass::L-smooth} \ref{pix_ass::variance} \ref{pix_ass::bias_corr} hold, consider the scheme in Algorithm~\ref{equ:alldisc}, and $\bone,\btwo \in (0,1)$, and $\btwo>\bone$, and $\epsilon, \lambda > 0$. 
$\|\lambda x_0\|_{\infty} \leq 1$.

We have
\bb
\frac{1}{T}\sum_{t=1}^{T} \E\mathcal{S}(x_t) \leq \frac{f(x_0) - f^*}{T\epsilon} + \frac{2D \beta_1 \beta_2 \sqrt{d}\|\dd f(x_0)\|}{T(1-\beta_2)}  + \frac{4\beta_1 L \epsilon d }{1-\beta_2} +\frac{2\sqrt{d}\sigma(1+\sqrt{C}) + 2\rho}{\sqrt{N}}+ 2L\epsilon d,
\ee
where $C = \beta_1^2 (1-\beta_2)\frac{1}{1+\beta_2} + (1-\beta_1)^2$, $D = \max\{1, \sigma/\left(2\sqrt{d}\beta_1 \beta_2^T\|\dd f(x_0)\|\right)\}$, and
\bb
\rho_t[k] =
    \begin{cases}
      0 &\text{if $\E [\sign(\tilde m_{t+1}^i[k])]=0$,}\\
      \E [\tilde m_{t+1}^i[k]] / \E [\sign(\tilde m_{t+1}^i[k])] &\text{else.}
    \end{cases}  
\ee
\end{thm}
\begin{proof}
Following Theorem \ref{pix_thm:phsI} from phase 1, 
once we have $\norm{\lambda x_0}_\infty\leq1$, 
 we stay within the constraint set with $\norm{\lambda x_t}\leq 1$ for all subsequent time $t\geq 0$.

For notation, write $\tilde M_{t+1} = \sum_{i=1}^N \sign(\tilde m^i_{t+1})$. This yields 
$x_{t+1} = x_{t} - \epsilon \sign(\tilde M_{t+1}) - \epsilon \lambda x_t$.
We have  
\bbb \label{equ::difference}
f(x_{t+1}) - f(x_t) &\leq \langle \dd f(x_t), x_{t+1} - x_t\rangle + \frac{L}{2}\|x_{t+1} - x_t\|_2^2 \ant{$L$-smoothness of $f$} \nonumber\\
&= -\epsilon \langle\dd f(x_t), \sign(\tilde M_{t+1}) + \lambda x_{t}\rangle + \frac{L}{2}\|x_{t+1} - x_t\|_2^2 \nonumber \\ 
&= -\epsilon \langle\dd f(x_t), \sign(\dd f(x_t)) + \lambda x_{t}\rangle  + \frac{L}{2}\|x_{t+1} - x_t\|_2^2  \nonumber\\ 
&\quad +\epsilon \langle \dd f(x_t),  \sign(\dd f(x_t)) - \sign(\tilde M_{t+1}))\rangle\nonumber \\
&\leq -\epsilon \mathcal{S}(x_t)  + 2L\epsilon^2 d+\epsilon \langle \dd f(x_t), \sign(\dd f(x_t)) - \sign(\tilde M_{t+1})\rangle, 
\eee
where we used $\norm{x_{t+1} - x_t}^2 = \epsilon^2\norm{\sign(\tilde M_{t+1}) + \lambda x_t}^2 \leq 4\epsilon^2 d$, because $\norm{\lambda x_t}_\infty \leq 1$.
% \qq{why $\norm{x_{t+1} - x_t}^2\leq 2\epsilon^2 d$? $x_{t}$ may not be bounded? Is it $1/2$ rather than $2$? }
% \lzchen{$\norm{x_{t+1} - x_t}^2 = \epsilon^2\norm{\sign(\tilde M_{t+1}) -\lambda x_t}^2 \leq 4\epsilon^2 d $ }

By Assumption~\ref{pix_ass::variance}, $\tilde m_{t+1}^1, \tilde m_{t+1}^2, \cdots, \tilde m_{t+1}^N$ are i.i.d., so $\E [\tilde m_{t+1}^i]$ and $ \E [\sign(\tilde m_{t+1}^i)]$ don't depend on $i$.
Hence we can define $R_{t+1} = \E [\tilde m_{t+1}^i] / \E [\sign(\tilde m_{t+1}^i)]$, where the division operation is element wise, so $R_{t+1} \in \R^d$.

By Assumption ~\ref{ass::bias_corr}, $R_t$ is non-negative, one special case for the ratio $R_t$ is when $\E[\sign(\tilde m_t^i[k])] = 0$, yet $\E[\tilde m_t^i[k]] \neq 0$, leading to $R_t[k] = +\infty$ for $k \in [d]$. In such instance, $P(\tilde m_t^i[k] > 0) = 1/2$ derived from the equation $\E[\sign(\tilde m_t^i[k])] = 2P(\tilde m_t^i[k] > 0) -1 = 0$, for $k \in [d]$. 

First, recognizing that $\E [\sign(\tilde M_t[k])] = 0$ is straightforward as we model it as a binomial distribution with success probability $p = 1/2$ for $t>0$.
This leads to the result $\E \dd f(x_t)[k] \left(\sign(\dd f(x_t)[k]) - \sign(\tilde M_{t}[k])\right) = \E \abs{\dd f(x_t)[k]}$.

Given that $\E[X] = \argmin_{z} \E \norm{X-z}_2$ defines the expectation of a random variable $X$ as the value $z$ minimizes the expected euclidean distance to $X$, and the $\textit{median}~X = \argmin_{z} \E \norm{X-z}_1$ defines the median as the value $z$ minimizing the expected absolute distance to $X$, for a R.V. $X$ in $\R$, recall our case where $P(\tilde m_t^i[k] > 0) = 1/2$, which is equivalent to that the median is $0$. From this, it follows that
\bb
\E \abs{\dd f(x_t)[k]} \leq \E [\E_\xi [\abs{\dd f(x_t;\xi_t^i)[k] - \dd f(x_t)[k]}_1]] \leq \E\sqrt{\E_\xi \norm{\dd f(x_t;\xi_t^i)[k] - \dd f(x_t)[k]}_2^2} \leq \sigma.
\ee
To bound the last term in \eqref{equ::difference} $\langle \dd f(x_t), \sign(\dd f(x_t)) - \sign(\tilde M_{t+1})\rangle $, we follow a structured approach. Here's an outline for bounding this term:

To bound the last term in Equation~\eqref{equ::difference}, $\langle \nabla f(x_t), \sign(\nabla f(x_t)) - \sign(\tilde M_{t+1})\rangle$, we follow a structured approach:

\begin{enumerate}
    \item \textbf{Transform Inner Product into Norm of Difference}: Using Lemma~\ref{lem:xylemma} to convert the inner product $\langle \dd f(x_t), \sign(\dd f(x_t)) - \sign(\tilde M_{t+1})\rangle $ into the norm of a difference.
    
    \item \textbf{Introduce $R_t$ as a De-bias Ratio}: $R_t$ is defined to adjust or correct for any bias in the expected value of $\tilde m_t^i$ and the expected sign of $\tilde m_t^i$ as in Assumption~\ref{pix_ass::bias_corr}.
    \item \textbf{Handle Cases of $R_t$ Separately}: Given the possibility of $R_t[k] = +\infty$, it's essential to treat the scenarios of $R_t[k] < +\infty$ and $R_t[k] = +\infty$ with separate proofs.
    \begin{itemize}
        \item For $R_t[k] < +\infty$, standard bounding techniques can be applied, potentially leveraging properties of $R_t$ to establish a finite upper bound.
        \item For $R_t[k] = +\infty$, it's actually bounding $\norm{\dd f(x_t)}$. This can be bounded by the variance of the stochastic gradient $g_t^i$.
    \end{itemize}
    \item \textbf{Merge Cases with Finite $\rho_t$ Replacing $R_t$}: After separately proving bounds for each case of $R_t$, the results are unified by substituting $R_t$ with a finite $\rho_t$, where $\rho_t$ serves a similar purpose but ensures a manageable, finite adjustment.
\end{enumerate}

% \qq{the decision of putting which part into a lemma seems arbitrary. The overall derivation below is still complicate.}

\textbf{Case I (Finite $R_{t+1}$)}

The first step is to expand this inner product, we have
\bb
&\E \langle \dd f(x_t), \sign(\dd f(x_t)) - \sign(\tilde M_{t+1})\rangle \\
&= \E \langle \dd f(x_t), \sign(\dd f(x_t)) - \sign(\frac{1}{N}\tilde M_{t+1})\rangle \\
&= \E \sum_{k=1}^d\dd f(x_t)[k]\left(\sign(\dd f(x_t)[k]) - \sign(\frac{1}{N}\tilde M_{t+1}[k])\right)\\
&= 2\E \sum_{k=1}^d R_{t+1}[k]\abs{\dd f(x_t)[k]/R_{t+1}[k] - \frac{1}{N}\tilde M_{t+1}[k]}\\
&= 2\E \sum_{k=1}^d R_{t+1}[k]\abs{\dd f(x_t)[k]/R_{t+1}[k] - \frac{1}{N}\sum_{i=1}^N\sign(\tilde m_{t+1}^i[k])}.\ant{Lemma~\ref{lem:xylemma} and Assumption~\ref{ass::bias_corr}}
\ee
By definition of $R_t$, it is a debiasing ratio between $\E [\tilde m_{t+1}^i]$ and $\E [\sign(\tilde m_{t+1}^i)]$, so we construct a difference between $\frac{1}{N}\sum_{i=1}^N \sign(\tilde m_{t+1}^i[k])$ and $\frac{1}{N}\sum_{i=1}^N\tilde m_{t+1}^i[k]$ by decoupling the difference between $\dd f(x_t)[k]/R_{t+1}[k]$ and $\frac{1}{N}\sign(\tilde m_{t+1}^i[k])$.
\bb
&\E R_{t+1}[k]\abs{\dd f(x_t)[k]/R_{t+1}[k] - \frac{1}{N}\sum_{i=1}^N\sign(\tilde m_{t+1}^i[k])} \\
&=\E R_{t+1}[k]\abs{\dd f(x_t)[k]/R_{t+1}[k] -\frac{1}{N}\sum_{i=1}^N\tilde m_{t+1}^i[k]/R_{t+1}[k] + \frac{1}{N}\sum_{i=1}^N\tilde m_{t+1}^i[k]//R_{t+1}[k]- \frac{1}{N}\sum_{i=1}^N\sign(\tilde m_{t+1}^i[k])} \\
&=\E R_{t+1}[k]\abs{\dd f(x_t)[k]/R_{t+1}[k] -\frac{1}{N}\sum_{i=1}^N\tilde m_{t+1}^i[k]/R_{t+1}[k]} + R_{t+1}[k]\abs{\frac{1}{N}\sum_{i=1}^N\tilde m_{t+1}^i[k]/R_{t+1}[k]- \frac{1}{N}\sum_{i=1}^N\sign(\tilde m_{t+1}^i[k])}  \\
&=\E\abs{\dd f(x_t)[k] -\frac{1}{N}\sum_{i=1}^N\tilde m_{t+1}^i[k]} + R_{t+1}[k]\abs{\frac{1}{N}\sum_{i=1}^N\tilde m_{t+1}^i[k]/R_{t+1}[k]- \frac{1}{N}\sum_{i=1}^N\sign(\tilde m_{t+1}^i[k])}.
\ee
The first term $\E\abs{\dd f(x_t)[k] -\frac{1}{N}\sum_{i=1}^N\tilde m_{t+1}^i[k]}$ doesn't depend on $R_{t+1}$, we can bound this term across $d$ coordinates using Lemma~\ref{lem:core}:
\bb
\E\sum_{k=1}^d\abs{\dd f(x_t)[k] -\frac{1}{N}\sum_{i=1}^N\tilde m_{t+1}^i[k]} &\leq \sqrt{d} \E \norm{\dd f(x_t) -\frac{1}{N}\sum_{i=1}^N\tilde m_{t+1}^i} \\
&\leq \sqrt{d} \E \norm{\dd f(x_t) -\frac{1}{N}\sum_{i=1}^N \left(\bone m_t^i+(1-\bone)g_t^i\right)} \\
&\leq \sqrt{d} \E \norm{\frac{1}{N}\sum_{i=1}^N \bone\left(\dd f(x_t) - m_t^i\right)}+\norm{\frac{1}{N}\sum_{i=1}^N(1-\bone)\left(\dd f(x_t) - g_t^i\right)} \\
&\leq \sqrt{d}\beta_1 \left(\beta_2^t\|\dd f(x_0)\| + \frac{2L\epsilon\sqrt{d}}{1-\beta_2}+ \frac{\sigma}{\sqrt{N(1+\btwo)}}\right) +  \frac{\sqrt{d}\sigma (1-\bone)}{\sqrt{N}}. \ant{Lemma~\ref{lem:core}}
\ee

The second term $\E R_{t+1}[k]\abs{\frac{1}{N}\sum_{i=1}^N\tilde m_{t+1}^i[k]/R_{t+1}[k]- \frac{1}{N}\sum_{i=1}^N\sign(\tilde m_{t+1}^i[k])}$ can be decoupled into the variance of $\frac{1}{N}\sum_{i=1}^N\sign(\tilde m_{t+1}^i[k])$ and the variance of $\frac{1}{N}\sum_{i=1}^N\tilde m_{t+1}^i[k]$:
\bb
&\E \sum_{k=1}^d R_{t+1}[k]\abs{\frac{1}{N}\sum_{i=1}^N\tilde m_{t+1}^i[k]/R_{t+1}[k]- \frac{1}{N}\sum_{i=1}^N\sign(\tilde m_{t+1}^i[k])}\\
&=\E \sum_{k=1}^d R_{t+1}[k]\abs{\frac{1}{N}\sum_{i=1}^N\tilde m_{t+1}^i[k]/R_{t+1}[k]-\E \tilde m_{t+1}^i[k]/R_{t+1}[k] + \E \tilde m_{t+1}^i[k]/R_{t+1}[k]- \frac{1}{N}\sum_{i=1}^N\sign(\tilde m_{t+1}^i[k])}\\
&=\E \sum_{k=1}^d R_{t+1}[k]\abs{\frac{1}{N}\sum_{i=1}^N\tilde m_{t+1}^i[k]/R_{t+1}[k]-\E \tilde m_{t+1}^i[k]/R_{t+1}[k] + \E \sign(\tilde m_{t+1}^i[k])- \frac{1}{N}\sum_{i=1}^N\sign(\tilde m_{t+1}^i[k])}\\
&=\E \sum_{k=1}^d R_{t+1}[k]\abs{\frac{1}{N}\sum_{i=1}^N\tilde m_{t+1}^i[k]/R_{t+1}[k]-\E \tilde m_{t+1}^i[k]/R_{t+1}[k]} + R_{t+1}[k]\abs{\E \sign(\tilde m_{t+1}^i[k])- \frac{1}{N}\sum_{i=1}^N\sign(\tilde m_{t+1}^i[k])}\\
&=\E \sum_{k=1}^d \abs{\frac{1}{N}\sum_{i=1}^N\tilde m_{t+1}^i[k]-\E \tilde m_{t+1}^i[k]} + R_{t+1}[k]\abs{\frac{1}{N}\sum_{i=1}^N\sign(\tilde m_{t+1}^i)-\E \sign(\tilde m_{t+1}^i)}\\
&\leq \E \sqrt{d}\norm{\frac{1}{N}\sum_{i=1}^N\tilde m_{t+1}^i-\E \tilde m_{t+1}^i} + \norm{R_{t+1}}\norm{\frac{1}{N}\sum_{i=1}^N\sign(\tilde m_{t+1}^i)-\E \sign(\tilde m_{t+1}^i)}.
\ee
Now we have got the variance of $\frac{1}{N}\sum_{i=1}^N\sign(\tilde m_{t+1}^i[k])$ and the variance of $\frac{1}{N}\sum_{i=1}^N\tilde m_{t+1}^i[k]$, let us bound them one by one: 
\paragraph{The variance of $\frac{1}{N}\sum_{i=1}^N\sign(\tilde m_{t+1}^i[k])$}
\bb
\sqrt{d}\E \norm{\frac{1}{N}\sum_{i=1}^N\tilde m_{t+1}^i-\E \tilde m_{t+1}^i}
&\leq \sqrt{d}\sqrt{\E \norm{\frac{1}{N}\sum_{i=1}^N\tilde m_{t+1}^i-\E \tilde m_{t+1}^i}^2}\\
&=\sqrt{d}\sqrt{\frac{1}{N^2}\sum_{i=1}^N\E\norm{\tilde m_{t+1}^i-\E \tilde m_{t+1}^i}^2}\\
&\leq \sqrt{\frac{Cd\sigma^2}{N}}, \ant{Lemma \ref{lem:martingale}}
\ee
where $C = \beta_1^2 (1-\beta_2)\frac{1}{1+\beta_2} + (1-\beta_1)^2$.
\paragraph{The variance of $\frac{1}{N}\sum_{i=1}^N\tilde m_{t+1}^i[k]$}
\bb
\norm{R_{t+1}}\E \norm{\frac{1}{N}\sum_{i=1}^N\sign(\tilde m_{t+1}^i)-\E \sign(\tilde m_{t+1}^i)} &\leq \sqrt{\E \norm{\sum_{i=1}^N \sign(\tilde m^i_{t+1}) / N - \E [\sign(\tilde m^i_{t+1})]}^2} \\
&= \norm{R_{t+1}}\sqrt{\frac{1}{N^2}\sum_{i=1}^N\E \norm{\sign(\tilde m^i_{t+1}) - \E [\sign(\tilde m^i_{t+1})]}^2} \\
&\leq \norm{R_{t+1}}\sqrt{\frac{1}{N}}.\ant{Lemma \ref{lem::sign_exp}}
\ee 
In above, we have the bound of the last term in \eqref{equ::difference} $\langle \dd f(x_t), \sign(\dd f(x_t)) - \sign(\tilde M_{t+1})\rangle$: 
\bb
&\E \langle \dd f(x_t), \sign(\dd f(x_t)) - \sign(\tilde M_{t+1})\rangle \\
&\leq 2\E\sum_{k=1}^d\abs{\dd f(x_t)[k] -\frac{1}{N}\sum_{i=1}^N\tilde m_{t+1}^i[k]} + 2\E \sum_{k=1}^d R_{t+1}[k]\abs{\frac{1}{N}\sum_{i=1}^N\tilde m_{t+1}^i[k]/R_{t+1}[k]- \frac{1}{N}\sum_{i=1}^N\sign(\tilde m_{t+1}^i[k])}\\
&\leq 2\sqrt{d} \E \norm{\dd f(x_t) -\frac{1}{N}\sum_{i=1}^N\tilde m_{t+1}^i} + 2\E \sqrt{d}\norm{\frac{1}{N}\sum_{i=1}^N\tilde m_{t+1}^i-\E \tilde m_{t+1}^i} + 2\norm{R_{t+1}}\norm{\frac{1}{N}\sum_{i=1}^N\sign(\tilde m_{t+1}^i)-\E \sign(\tilde m_{t+1}^i)}\\
&\leq 2\sqrt{d}\beta_1 \left(\beta_2^t\|\dd f(x_0)\| + \frac{2L\epsilon\sqrt{d}}{1-\beta_2}+ \frac{\sigma}{\sqrt{N(1+\btwo)}}\right) +  2\frac{\sqrt{d}\sigma (1-\bone)}{\sqrt{N}} + 2\sqrt{\frac{Cd\sigma^2}{N}} + 2\norm{R_{t+1}}\sqrt{\frac{1}{N}}.
\ee
\textbf{Case II (Infinite $R$)} 

From our discussion above, we know that $P(\tilde m_t^i[k] > 0) = 1/2$ since $\E[\sign(\tilde m_t^i[k])] = 2P(\tilde m_t^i[k] > 0) -1 = 0$, where $k \in [d]$. For notion, write $\mathcal{D} = \{j\in [d] ~|~\E [\sign(\tilde m_{t+1}^i[j])]=0\}$. In this case, we have
\bb
\E \sum_{j\in \mathcal{D}}\dd f(x_t)[j] \left(\sign(\dd f(x_t)[j]) - \sign(\tilde M_{t}[j])\right) &= \E \sum_{j\in \mathcal{D}}\abs{\dd f(x_t)[j]}\\
&\leq \E \left[\E_\xi \sum_{j\in \mathcal{D}}\abs{\dd f(x_t;\xi_t^i)[j] - \dd f(x_t)[j]}\right] \\
&\leq \E\sqrt{\E_\xi \sum_{j\in \mathcal{D}}\norm{\dd f(x_t;\xi_t^i)[j] - \dd f(x_t)[j]}_2^2} \\
&\leq \sigma.
\ee
So, the inner product $\langle \dd f(x_t), \sign(\dd f(x_t)) - \sign(\tilde M_{t+1})\rangle$ is still bounded. Hence we can merge both cases into a unified bound by simply replacing $R_t$ by $\rho_t$: 
\bb
\rho_t[k] =
    \begin{cases}
      0 &\text{if $\E [\sign(\tilde m_{t+1}^i[k])]=0$,}\\
      \E [\tilde m_{t+1}^i[k]] / \E [\sign(\tilde m_{t+1}^i[k])] &\text{else.}
    \end{cases}  
\ee
Adding one constant $D\geq1$ to make the bound in finite case adpative to infinite case:
\bb
\sigma \leq 2D\sqrt{d}\beta_1 \beta_2^t\|\dd f(x_0)\|, \forall t, 1\leq t\leq T.
\ee
Hence,
\bb
&\E \sum_{j\in \mathcal{D}}\dd f(x_t)[j] \left(\sign(\dd f(x_t)[j]) - \sign(\tilde M_{t}[j])\right)\\
&\leq 2D\sqrt{d}\beta_1\beta_2^t\|\dd f(x_0)\| + \frac{4Ld\bone\epsilon}{1-\btwo} + \frac{2\sqrt{d}\sigma(1+\sqrt{C}) + 2\norm{\rho_{t+1}}}{\sqrt{N}}. 
\ee
Finally, we have the bound for both cases:
\bb
&\E \langle \dd f(x_t), \sign(\dd f(x_t)) - \sign(\tilde M_{t+1})\rangle \\
&\leq 2\sqrt{d}\beta_1 \left(\beta_2^t\|\dd f(x_0)\| + \frac{2L\epsilon\sqrt{d}}{1-\beta_2}+ \frac{\sigma}{\sqrt{N(1+\btwo)}}\right) +  2\frac{\sqrt{d}\sigma (1-\bone)}{\sqrt{N}} + 2\sqrt{\frac{Cd\sigma^2}{N}} + 2\norm{\rho_{t+1}}\sqrt{\frac{1}{N}}\\
&\leq 2D\sqrt{d}\beta_1\beta_2^t\|\dd f(x_0)\| + \frac{4Ld\bone\epsilon}{1-\btwo} + \frac{2\sqrt{d}\sigma(1+\sqrt{C}) + 2\norm{\rho_{t+1}}}{\sqrt{N}}. 
\ee

% \bb
% &\leq 2 \sqrt{d} R_{t+1} \E \| \dd f(x_t)/R_{t+1} - \frac{1}{N}\tilde M_{t+1}\| {\ant{Lemma~\ref{lem:diff_sign}}} \\
% &= 2 \sqrt{d} R_{t+1} \E \|\sum_{i=1}^N \sign(\tilde m^i_{t+1}) / N - \dd f(x_t)/R_{t+1}\|  \ant{definition of $\tilde M_{t+1}$}\\ 
% %\ant{by \ref{lem:diff_sign}} \\
% &\leq 2\sqrt{d} R_{t+1}\sqrt{\frac{1+C\sigma^2/R_{t+1}^2}{N}} + 2 \sqrt{d} R_{t+1} \E \left\|\sum_{i=1}^N \tilde m^i_{t+1}/(NR_{t+1}) - \dd f(x_t) / R_{t+1}\right\| \ant{Lemma \ref{lem::diff_momentum}}\\
% &\leq 2\sqrt{d} \sqrt{\frac{R_{t+1}^2+C\sigma^2}{N}} + 2 \sqrt{d} R_{t+1} \E \left\|\frac{1}{NR_{t+1}}\sum_{i=1}^N \beta_1 (m_t^i - \dd f(x_t))\right\| + 2 \sqrt{d} R_{t+1}\E\left\| \frac{1}{NR_{t+1}}\sum_{i=1}^T(1-\beta_1)\left(g_t^i-\dd f(x_t)\right) \right\| \\
% &\leq 2\sqrt{d} \sqrt{\frac{R_{t+1}^2+C\sigma^2}{N}} + 2 \sqrt{d} R_{t+1} \frac{1}{NR_{t+1}} N \beta_1 \left(\beta_2^t\|\dd f(x_0)\| + \frac{2L\epsilon\sqrt{d}}{1-\beta_2} + \frac{\sigma}{\sqrt{N(1+\beta_2)}}\right) + 2 \sqrt{d} \frac{\sigma (1-\beta_1)}{\sqrt{N}} \ant{Lemma \ref{lem:core}}\\
% &\leq 2\sqrt{d} \sqrt{\frac{R_{t+1}^2+C\sigma^2}{N}} + 2 \sqrt{d}\beta_1 \left(\beta_2^t\|\dd f(x_0)\| + \frac{2L\epsilon\sqrt{d}}{1-\beta_2}\right) +  \frac{2\sqrt{d}\sigma}{\sqrt{N}},
% \ee
% where $C = \beta_1^2 (1-\beta_2)\frac{1}{1+\beta_2} + (1-\beta_1)^2$.

Then we have
\bb
f(x_{t+1}) - f(x_t) &\leq  -\epsilon \mathcal{S}(x_t)  + 2L\epsilon^2 d+\epsilon \langle \dd f(x_t), \sign(\dd f(x_t)) - \sign(\tilde M_{t+1})\rangle \\
& \leq  -\epsilon \mathcal{S}(x_t)  + 2L\epsilon^2 d+\epsilon \left(2D\sqrt{d}\beta_1\beta_2^t\|\dd f(x_0)\| + \frac{4Ld\bone\epsilon}{1-\btwo} + \frac{2\sqrt{d}\sigma(1+\sqrt{C}) + 2\norm{\rho_{t+1}}}{\sqrt{N}} \right),
\ee
Hence, a telescope yields
\bb
\frac{1}{T}\sum_{t=1}^{T} \E\mathcal{S}(x_t) \leq \frac{f(x_0) - f^*}{T\epsilon} + \frac{2D\beta_1 \beta_2 \sqrt{d}\|\dd f(x_0)\|}{T(1-\beta_2)}  + \frac{4\beta_1 L \epsilon d }{1-\beta_2} +\frac{2\sqrt{d}\sigma(1+\sqrt{C}) + 2\rho}{\sqrt{N}}+ 2L\epsilon d,
\ee
where $\rho = \max_{1\leq t\leq T} \norm{\rho_t}$.
\end{proof}

\begin{lem}\label{lem:diff_sign}
    Let $(X,Y)$ is a joint random variable on $\RR^d \times \RR^d$. For any constant $a \in (0, +\infty)$, we have $$\E [\langle X, \sign(X) - \sign(Y) \rangle] \leq 2 a \sqrt{d}\E \|X/a-Y\|.$$
\end{lem} 
%\qq{No proof for Lemma ~\ref{lem:diff_sign}?? Is it taken from a previous paper? If so, cite.}\lzchen{i added proof here.}
\begin{proof}
Without loss of generality, set $a = 1$. 
\bb
    \E [\langle X, \sign(X) - \sign(Y) \rangle] &= \E[\norm{X}_1 - \langle X, \sign(Y)\rangle] \\
    &\leq 2\E[\norm{X-Y}_1] \red{\ant{Lemma~\ref{lem:xylemma}}} \\
    &\leq 2\sqrt{d}\E[\norm{X-Y}] \ant{by Cauchy-Schwarz,}
\ee
where $\norm{\cdot}_1$ is the $\ell_1$ norm and $\norm{\cdot}$ denotes the Euclidean norm.
\end{proof}
\begin{lem}\label{lem:xylemma}
For any $x, y\in \RR$, we have 
$$\abs{x} - x \sign(y) \leq 2\abs{x- y}.$$
\end{lem}
\begin{proof}
If $\sign(y) = \sign(x)$,  we have 
$\abs{x} - x\sign(y) =0 \leq 2\abs{x-y}$. 

If $\sign(y) = - \sign(x),$ we have 
$\abs{x} - x \sign(y) =2\abs{x}\leq 2 \abs{x} + 2\abs{y} = 2 \abs{x-y}$. 

If $\sign(y) = 0$, we have $\abs{x} - x\sign(y) = \abs{x} = \abs{x-y}\leq 2 \abs{x-y}.$
\end{proof}

\begin{lem}\label{lem::sign_exp}
Let $X$ be a random variable in $\R$, we have $\E\norm{\sign(X)-\E[\sign(X)]}^2 < 1$.
\end{lem}
\begin{proof}
The result is a direct derivation from Bernoulli distribution's variance,
$$\E\norm{\sign(X)-\E[\sign(X)]}^2 = \E[\sign(X)^2] - \E[\sign(X)]^2 < 1.$$
% \qq{should not the $\leq 1$ bound goes into the lemma? It is what actually used.}\lzchen{we actually used the property that it's less or than 1.}
\end{proof}
\begin{lem}\label{lem:core}
Following the same setting in Theorem~\ref{pix_thm:majority voting}, we have
    \bb 
    \|\frac{1}{N}\sum_{i=1}^N m_t^i - \dd f(x_t)\| \leq \beta_2^t\|\dd f(x_0)\| + \frac{2L\varepsilon\sqrt{d}}{1-\beta_2} + \frac{\sigma}{\sqrt{N(1+\beta_2)}}. 
    \ee 
\end{lem}
\begin{proof}
    
We use the notions: $g_t^i : = \dd f(x_t;\xi_t^i)$, $M_t = \frac{1}{N}\sum_{i=1}^N m_t^i$, $\varepsilon_t:= M_t  - \dd f(x_t)$, $\Bar{g_t} = \frac{1}{N}\sum_{i=1}^N g_t^i$, $\delta_t := \Bar{g_t} - \dd f(x_t)$, and $s_t = \dd f(x_{t-1}) - \dd f(x_t)$

\bb
\varepsilon_t &= M_t - \dd f(x_t)\\
&= \beta_2 M_{t-1} +(1-\beta_2)\Bar{g_t} - \dd f(x_t)\\
&= \beta_2 (M_{t-1} - \dd f(x_{t-1})) + (1-\beta_2) (\Bar{g_t} - \dd f(x_t)) + \beta_2 (\dd f(x_{t-1}) - \dd f(x_t) \\
&= \beta_2 \varepsilon_{t-1} + (1-\beta_2) \delta_t + \beta_2 s_t. \\
\ee
That is
$$
\varepsilon_t =\beta_2 \varepsilon_{t-1} + (1-\beta_2) \delta_t + \beta_2 s_t.
$$

% Note that the smoothness of the gradient implies: 
% $$
% \|s_t \| = \|\dd f(x_{t-1}) -\dd f(x_t)\| \qq{period}
% \qq{\text{is not this just the definition?}}
% $$

Under the $L$-smoothness assumption \ref{pdx_ass::L-smooth}:
\begin{align}
    \|s_t \| = \|\dd f(x_{t-1}) -\dd f(x_t)\| \leq L\|x_{t-1} - x_t\| \leq 2L \sqrt{d} \epsilon, \label{equ: strong}    
\end{align}
where $\varepsilon $ is the step size.
Using mathematical induction, we have
\begin{align}
    \varepsilon_t =\btwo^t \varepsilon_{0} +  \sum_{i=1}^t \btwo^{t-i+1} s_i + (1-\btwo) \sum_{i=1}^t \btwo^{t-i} \delta_t \label{equ: error}.
\end{align}
By taking the norms of both sides of the above equation and using the strong bound \ref{equ: strong} we obtain
\bb
\|\varepsilon_t\| \leq \btwo^t \|\varepsilon_0\|+ 2L\sqrt{d}\epsilon \sum_{i=1}^t \btwo^{t-i+1} + (1-\btwo) \|\sum_{i=1}^t \btwo^{t-i} \delta_t\|.
\ee
Taking expectations on both sides,
\bb
\E \|\varepsilon_t\| \leq \btwo^t \|\varepsilon_0\|+  \frac{2L\sqrt{d}\varepsilon}{1-\btwo} + (1-\btwo) \|\sum_{i=1}^t \btwo^{t-i} \delta_t\|. 
\ee

Note that r.v.s $(\delta_i)_{1\leq i \leq t}$ are mean zero, using \ref{lem:martingale}, we have

\bb
\E \left\|\sum_{i=1}^t \btwo^{t-i} \delta_i\right\| = \sqrt{\E \sum_{i=1}^t \btwo^{2t-2i} \frac{\sigma^2}{N}}\leq \frac{\sigma}{\sqrt{N(1-\btwo^2)}}
\ee

Hence,
\bb
\E \|\varepsilon_t\| \leq \btwo^t \|\varepsilon_0\|+  \frac{2 L\sqrt{d}\varepsilon}{1-\btwo} + \frac{\sigma}{\sqrt{N(1+\btwo)}}.
\ee
Note that $M_0 = 0$ under our setting, so $\varepsilon_0 = -\dd f(x_0)$, we have
\bb
\E \|\varepsilon_t\| \leq \btwo^t \|\dd f(x_0)\|+  \frac{2 L\sqrt{d}\varepsilon}{1-\btwo} + \frac{\sigma}{\sqrt{N(1+\btwo)}}.
\ee
\end{proof}

\begin{lem}[Cumulative error of stochastic gradient~\cite{bernstein_signsgd_2018}]\label{lem:martingale}
Assume the same settings as in 
Theorem~\ref{pix_thm:majority voting}. %we have that 
Define $Y_k \defeq \sum_{l=1}^k \alpha_\ell \delta_l$ where $\delta_t :=  \Bar{g_t} - \dd f(x_t)$ with $\Bar{g_t}  = \sum_{i=1}^N g_t^i$ and  
$g_t^i : = \dd f(x_t;\xi_t^i)$ following the update  in  \eqref{equ:alldisc}, 
and $\{\alpha_\ell\colon \ell = 0, 1,\ldots\}$ is a deterministic sequence. 
Then $Y_k$ is a martingale,  and 
	$$
	\E\left[\left[\sum_{l=1}^k  \alpha_l  \delta_l\right]^2\right] =  \frac{1}{N}\sum_{l=1}^k \alpha_l^2 {  \sigma}^2. $$
 % \red{double check equality or inequality}\lzchen{it's an equality.}
\end{lem}
\begin{proof}
	We simply check the definition of martingales. 
	First, we have 
	\begin{align*}
	\E[|Y_k|] &= \E\left[\abs{\sum_{l=1}^k\alpha_l \delta_l}\right] \\  
    &\leq \sum_l  |\alpha_l|\E[|\delta_l|] \ant{triangle inequality}\\
    &= \sum_l  |\alpha_l|  \E[\E[|\delta_l| | x_l]] \ant{law of total probability}\\
   &\leq \sum_l  |\alpha_l|\E[\sqrt{\E[ \delta_l^2  |x_l]} ] \ant{Jensen's inequality}\\
   &\leq \sum_l |\alpha_l|   \sigma  <\infty  \ant{Assumption~\ref{pix_ass::variance}.}
	\end{align*}
	Second, again using the law of total probability,
	\begin{align*}
	\E[Y_{k+1} | Y_1,...,Y_k]  &= \E\left[\sum_{l=1}^{k+1}\alpha_l \delta_l  \middle|  \alpha_1 \delta_1, ...,  \alpha_k \delta_k \right]  \\
	&=  Y_k  +  \alpha_{k+1}\E\left[ \delta_{k+1}  \middle|  \alpha_1 \delta_1, ...,  \alpha_k \delta_k  \right]\\
	&= Y_k  +   \alpha_{k+1}\E\left[  \E\left[ \delta_{k+1}  \middle|  x_{k+1},  \alpha_1 \delta_1, ...,  \alpha_k \delta_k  \right] | \alpha_1 \delta_1, ...,  \alpha_k \delta_k \right]\\
	&= Y_k  + \alpha_{k+1}\E\left[  \E\left[ \delta_{k+1}  \middle|  x_{k+1}\right] | \alpha_1 \delta_1, ...,  \alpha_k \delta_k \right]\\
	&=Y_k.
	\end{align*}
	This completes the proof that it is a martingale. We now make use of the properties of martingale difference sequences to establish a variance bound on the martingale.
	\begin{align*}
	\E[[\sum_{l=1}^k \alpha_l \delta_l ]^2 ]  &=  \sum_{l=1}^k \E[  \alpha_l^2 \delta_l^2]   +  2\sum_{l<j} \E[\alpha_l\alpha_j \delta_l \delta_j]\\
	&=\sum_{l=1}^k \alpha_l^2  \E[\E[\delta_l^2| \delta_1,...,\delta_{l-1} ]]   +   2\sum_{l<j} \alpha_l\alpha_j  \E\Big[\delta_l \E\big[ \E[\delta_j | \delta_1,...,\delta_{j-1}] \big| \delta_l \big]\Big]\\
    &=\sum_{l=1}^k \alpha_l^2  \E[\E[\E[\delta_l^2| x_l,\delta_1,...,\delta_{l-1} ]| \delta_1,...,\delta_{l-1} ]]   +   0\\
	&=\frac{1}{N}\sum_{l=1}^k \alpha_l^2   \sigma^2.
	\end{align*}
\end{proof} %The consequence of this lemma is that we are able to treat $\delta_1,...,\delta_k$ as if they are  independent, even though they are not---clearly $\delta_l$ is dependent on $\delta_1,...,\delta_{l-1}$ through $x_l$. 
As a direct result of Lemma \ref{lem:martingale}, we have the following. 
%is lemma \ref{lem::var_mom}.
\begin{lem}\label{lem::var_mom}
    %Adhering to the configurations outlined in 
    Under the same settings as in 
    Theorem \ref{thm:majority voting}, 
    we have 
    %e proceed by stating 
    $$
    \E\norm{\tilde m^i_{t+1} - \E [\tilde m^i_{t+1}]}^2 \leq \left(\beta_1^2 (1-\beta_2)\frac{1}{1+\beta_2} + (1-\beta_1)^2\right)\sigma^2. 
    $$
\end{lem}
\begin{proof}
    \bb
    \tilde m_{t+1}^i &= \beta_1 m_t^i + (1-\beta_1) g_t^i\\
    &=\beta_1 (1-\beta_2)\left(g_{t-1}^i + \beta_2 g_{t-2}^i + \cdots + \beta_2^{t-1} g_{0}^i\right) + (1-\beta_1) g_t^i.
    \ee
    Note that 
    \bb
    \beta_1^2 (1-\beta_2)^2\left(1+ \beta_2^2 + \cdots + \beta_2^{2(t-1)}\right) +  (1-\beta_1)^2&= \beta_1^2 (1-\beta_2)^2\frac{1-\beta_2^{2t}}{1-\beta_2^2} +  (1-\beta_1)^2. 
    \ee
    By using lemma \ref{lem:martingale}, we have  $$
    \E\norm{\tilde m^i_{t+1} - \E [\tilde m^i_{t+1}]}^2 \leq \left(\beta_1^2 (1-\beta_2)\frac{1}{1+\beta_2} + (1-\beta_1)^2\right)\sigma^2. 
    $$
\end{proof}
\subsubsection{Averaging Update Convergence}\label{sec::avg}
Assume $f\colon \RR^d\to \RR$ is $L$-smooth, $N$ is the number of workers, on the  $i$-th worker, consider the following scheme based on the averaging: 
\bbb \label{equ:avg} 
\begin{split}
& g_t^i : = \dd f(x_t;\xi_t^i),~~~~~~\forall i =1,\ldots, N\\
& m_{t+1}^i = \beta_2m_t^i + (1-\beta_2) g^i_t,~~~~~~\forall i =1,\ldots, N\\
& \tm_{t+1}^i = \beta_1 m_t^i + (1-\beta_1) g^i_t  ,~~~~~~\forall i =1,\ldots, N\\
& x_{t+1} = x_t - \epsilon \left(\frac{1}{N}\sum_{i=1}^N \sign(\tilde m^i_{t+1}) + \lambda x_{t} \right).  \ant{Average aggregation} \\
\end{split} 
\eee
\begin{thm}[Convergence in Phase II] \label{pix_thm:averaging}
Under Assumption \ref{pdx_ass::L-smooth} \ref{pix_ass::variance}, consider the scheme in~\eqref{equ:avg}
, and $\bone,\btwo \in (0,1)$, and $\btwo>\bone$, and $\epsilon, \lambda > 0$. $\|\lambda x_0\|_{\infty} \leq 1$.
We have
\bb
\frac{1}{T}\sum_{t=1}^{T} \E\mathcal{S}(x_t)\leq \frac{f(x_0) - f^*}{T\epsilon} + \frac{2\beta_1 \beta_2 \sqrt{d}\|\dd f(x_0)\|}{T(1-\beta_2)} + \frac{4\beta_1 L \epsilon d }{1-\beta_2} + \frac{2\beta_1 \sigma}{\sqrt{1+\beta_2}} + 2(1-\beta_1)\sigma+ 2L\epsilon d.
\ee
\end{thm}
\begin{proof}
For notation, write $\tilde M_{t+1} = \sum_{i=1}^N \sign(\tilde m^i_{t+1})$. This yields 
$x_{t+1} = x_{t} - \epsilon\tilde M_{t+1} - \epsilon \lambda x_t$. 

Following Theorem \ref{pix_thm:phsI} from phase 1, 
once we have $\norm{\lambda x_0}_\infty\leq1$, 
 we stay within the constraint set with $\norm{\lambda x_t}\leq 1$ for all subsequent time $t\geq 0$.

Following a similar procedure in ~\ref{pix_thm:majority voting}, we have
\bb 
f(x_{t+1}) - f(x_t) &\leq \langle \dd f(x_t), x_{t+1} - x_t\rangle + \frac{L}{2}\|x_{t+1} - x_t\|_2^2 \\
&\leq -\epsilon \langle\dd f(x_t), \tilde M_{t+1} + \lambda x_{t}\rangle + \frac{L}{2}\|x_{t+1} - x_t\|_2^2 \\ 
&\leq -\epsilon \langle\dd f(x_t), \sign(\dd f(x_t)) + \lambda x_{t}\rangle  + \frac{L}{2}\|x_{t+1} - x_t\|_2^2 \\ 
&\quad +\epsilon \langle \dd f(x_t),  \sign(\dd f(x_t)) - \tilde M_{t+1}\rangle \\
&\leq -\epsilon \mathcal{S}(x_t)  + 2L\epsilon^2 d+\epsilon \langle \dd f(x_t), \sign(\dd f(x_t)) - \tilde M_{t+1}\rangle.
\ee 
Let us bound the last term $\langle \dd f(x_t), \sign(\dd f(x_t)) - \tilde M_{t+1} \rangle$,
\bb
&\E \langle \dd f(x_t), \sign(\dd f(x_t)) - \tilde M_{t+1} \rangle \\
& = \E \langle \dd f(x_t), \sign(\dd f(x_t)) -  \frac{1}{N}\sum_{i=1}^N \sign(\tilde m^i_{t+1})\rangle \\
& = \sum_{i = 1}^N \frac{1}{N}\E \langle \dd f(x_t), \sign(\dd f(x_t)) -  \sign(\tilde m^i_{t+1})\rangle \\
& = \E \langle \dd f(x_t), \sign(\dd f(x_t)) -  \sign(\tilde m^i_{t+1})\rangle \ant{$\{\tilde m^i_{t+1}\}_{1\leq i \leq N}$ are independent} \\
&\leq 2\sqrt{d}\E \left\|\dd f(x_t) - \tilde m^i_{t+1}\right\| \ant{Lemma \ref{lem:diff_sign}} \\
& \leq 2\sqrt{d} \E \left[\beta_1\left\|\dd f(x_t) - m^i_{t}\right\| + (1-\beta_1)\left\|\dd f(x_t) - g^i_{t}\right\|\right] \ant{triangle inequality}\\
& \leq 2\sqrt{d} \left(\beta_1 \left(\beta_2^t\|\dd f(x_0)\| + \frac{2L\epsilon\sqrt{d}}{1-\beta_2} + \frac{\sigma}{\sqrt{1+\beta_2}}\right) + (1-\beta_1)\sigma\right). \ant{Lemma \ref{lem:core}}
\ee
Then we have
\bb
f(x_{t+1}) - f(x_t) &\leq -\epsilon \mathcal{S}(x_t)  + 2L\epsilon^2 d+\epsilon \langle \dd f(x_t), \sign(\dd f(x_t)) - \tilde M_{t+1}\rangle \\
& \leq -\epsilon \mathcal{S}(x_t)  + 2L\epsilon^2 d + 2\epsilon \sqrt{d} \left(\beta_1 \left(\beta_2^t\|\dd f(x_0)\| + \frac{2L\epsilon\sqrt{d}}{1-\beta_2} + \frac{\sigma}{\sqrt{1+\beta_2}}\right) + (1-\beta_1)\sigma\right).
\ee
Hence, a telescope yields
\bb
\frac{1}{T}\sum_{t=1}^{T} \E\mathcal{S}(x_t) \leq \frac{f(x_0) - f^*}{T\epsilon} + \frac{2\beta_1 \beta_2 \sqrt{d}\|\dd f(x_0)\|}{T(1-\beta_2)} + \frac{4\beta_1 L \epsilon d }{1-\beta_2} + \frac{2\beta_1 \sigma\sqrt{d}}{\sqrt{1+\beta_2}} + 2(1-\beta_1)\sqrt{d}\sigma+ 2L\epsilon d.
\ee
\end{proof}
\subsubsection{Global Lion Convergence}\label{sec::glob}
Assume $f\colon \RR^d\to \RR$ is $L$-smooth, $N$ is the number of workers, on the $i$-th worker,
consider the following scheme based on the global Lion: 
\bbb \label{equ:global} 
\begin{split}
& g_t^i : = \dd f(x_t;\xi_t^i)\\
& m_{t+1}^i = \beta_2m_t^i + (1-\beta_2) g^i_t\\
& \tm_{t+1}^i = \beta_1 m_t^i + (1-\beta_1) g^i_t  \\
& x_{t+1} = x_t - \epsilon \left( \sign(\frac{1}{N}\sum_{i=1}^N \tilde m^i_{t+1}) + \lambda x_{t} \right).  \ant{Global Lion} \\
\end{split} 
\eee 

\begin{thm}[Convergence in Phase II] \label{pix_thm:global}Under Assumption \ref{pdx_ass::L-smooth} and \ref{pix_ass::variance}, 
consider the scheme in ~\eqref{equ:global}
, and $\bone,\btwo \in (0,1)$, and $\btwo>\bone$, and $\epsilon, \lambda > 0$. $\|\lambda x_0\|_{\infty} \leq 1$.
We have
\bb
\frac{1}{T}\sum_{t=1}^{T} \E\mathcal{S}(x_t) \leq \frac{f(x_0) - f^*}{T\epsilon} + \frac{2\beta_1 \beta_2 \sqrt{d}\|\dd f(x_0)\|}{T(1-\beta_2)} + \frac{4\beta_1 L \epsilon d }{1-\beta_2} + \frac{2\sqrt{d} \sigma}{\sqrt{N}}.
\ee
\end{thm}
\begin{proof}For notation, write $\tilde G_{t+1} = \frac{1}{N}\sum_{i=1}^N \tilde m^i_{t+1}$. This yields 
$x_{t+1} = x_{t} - \epsilon\sign(\tilde G_{t+1}) - \epsilon \lambda x_t$. 

Following Theorem \ref{pix_thm:phsI} from phase 1, 
once we have $\norm{\lambda x_0}_\infty\leq1$, 
 we stay within the constraint set with $\norm{\lambda x_t}\leq 1$ for all subsequent time $t\geq 0$.

Following the same procedure in ~\ref{pix_thm:majority voting}, we have
\bb 
f(x_{t+1}) - f(x_t) &\leq \langle \dd f(x_t), x_{t+1} - x_t\rangle + \frac{L}{2}\|x_{t+1} - x_t\|_2^2 \\
&\leq -\epsilon \langle\dd f(x_t), \sign(\tilde G_{t+1}) + \lambda x_{t}\rangle + \frac{L}{2}\|x_{t+1} - x_t\|_2^2 \\ 
&\leq -\epsilon \langle\dd f(x_t), \sign(\dd f(x_t)) + \lambda x_{t}\rangle  + \frac{L}{2}\|x_{t+1} - x_t\|_2^2 \\ 
&\quad +\epsilon \langle \dd f(x_t),  \sign(\dd f(x_t)) - \sign(\tilde G_{t+1})\rangle \\
&\leq -\epsilon \mathcal{S}(x_t)  + 2L\epsilon^2 d+\epsilon \langle \dd f(x_t), \sign(\dd f(x_t)) - \sign(\tilde G_{t+1})\rangle.
\ee 
Let us bound $\langle \dd f(x_t), \sign(\dd f(x_t)) - \sign(\tilde G_{t+1}) \rangle$,
\bb
&\E \langle \dd f(x_t), \sign(\dd f(x_t)) - \sign(\tilde G_{t+1}) \rangle \\
& = \E \langle \dd f(x_t), \sign(\dd f(x_t)) -  \sign(\frac{1}{N}\sum_{i=1}^N \tilde m^i_{t+1})\rangle \\
&\leq 2\sqrt{d}\E \left\|\dd f(x_t) - \frac{1}{N}\sum_{i=1}^N\tilde m^i_{t+1}\right\| \ant{Lemma \ref{lem:diff_sign}} \\
& \leq 2\sqrt{d} \E \left[\beta_1\left\|\dd f(x_t) - \frac{1}{N}\sum_{i=1}^N m^i_{t}\right\| + (1-\beta_1)\left\|\dd f(x_t) - \frac{1}{N}\sum_{i=1}^N g^i_{t}\right\|\right] \ant{triangle inequality}\\
& \leq 2\sqrt{d} \left(\beta_1 \left(\beta_2^t\|\dd f(x_0)\| + \frac{2L\epsilon\sqrt{d}}{1-\beta_2} + \frac{\sigma}{\sqrt{N(1+\beta_2})}\right) + \frac{(1-\beta_1)\sigma}{\sqrt{N}}\right) \ant{Lemma \ref{lem:core}} \\
& \leq 2\sqrt{d} \left(\beta_1 \left(\beta_2^t\|\dd f(x_0)\| + \frac{2L\epsilon\sqrt{d}}{1-\beta_2}\right) + \frac{(1-\beta_1)\sigma}{\sqrt{N}}\right).
\ee
Then we have
\bb
f(x_{t+1}) - f(x_t) &\leq -\epsilon \mathcal{S}(x_t)  + 2L\epsilon^2 d+\epsilon \langle \dd f(x_t), \sign(\dd f(x_t)) - \tilde M_{t+1}\rangle \\
& \leq -\epsilon \mathcal{S}(x_t)  + 2L\epsilon^2 d + 2\epsilon \sqrt{d}\left(\beta_1 \left(\beta_2^t\|\dd f(x_0)\| + \frac{2L\epsilon\sqrt{d}}{1-\beta_2}\right) + \frac{(1-\beta_1)\sigma}{\sqrt{N}}\right).
\ee
Hence, a telescope yields
\bb
\frac{1}{T}\sum_{t=1}^{T} \E\mathcal{S}(x_t) \leq \frac{f(x_0) - f^*}{T\epsilon} + \frac{2\beta_1 \beta_2 \sqrt{d}\|\dd f(x_0)\|}{T(1-\beta_2)} + \frac{4\beta_1 L \epsilon d }{1-\beta_2} + \frac{2(1-\beta_1)\sqrt{d} \sigma}{\sqrt{N}} + 2L\epsilon d.
\ee
\end{proof}

\end{document}